\documentclass[12pt,preprint]{aastex}

\def\dw#1{}			
\def\jm#1{}

\begin{document}


\title{Small Scale Structure at High Redshift:
IV. Low Ionization Gas Intersecting Three Lines of Sight to 
Q2237+0305 \altaffilmark{1}\\
 }

\vskip 1.5cm

\author{Michael Rauch\altaffilmark{2}, Wallace L.W. Sargent\altaffilmark{3},
Thomas A. Barlow\altaffilmark{3}, Robert A. Simcoe\altaffilmark{3}}
\altaffiltext{1}{The observations were made at the W.M. Keck Observatory
which is operated as a scientific partnership between the California
Institute of Technology and the University of California; it was made
possible by the generous support of the W.M. Keck Foundation.}
\altaffiltext{2}{Carnegie Observatories, 813 Santa Barbara Street,
Pasadena, CA 91101, USA}
\altaffiltext{3}{Astronomy Department, California Institute of Technology,
Pasadena, CA 91125, USA}
\vfill

\begin{abstract}

We have obtained Keck HIRES spectra of three images of the quadruply
gravitationally lensed QSO 2237+0305 to study low ionization absorption
systems and their differences in terms of projected velocity and column
density across the lines of sight.  We detect CaII absorption from our Galaxy,
and a  system of High Velocity
Clouds from the lensing galaxy (z=0.039) with
multiple CaII components in all three sightlines. Unlike the situation in
our Galaxy there is no prominent CaII absorption component (with an
equivalent width exceeding 60-70 m\AA ) close to the velocity centroid of the
lensing galaxy 2237+0305. Instead, CaII components with  total
equivalent widths similar to those of Galactic intermediate and high
velocity clouds are spread out over several hundred kms$^{-1}$ in
projection along the sightlines at impact parameters of less than one
kpc through the bulge of the galaxy. A CaII absorbing thick disk like
in our Galaxy does not seem to extend into the bulge region of the
2237+0305 galaxy, whereas high velocity clouds seem to be a more
universal feature.  We have also studied three low ionization MgII-FeII
systems in detail.  All three MgII systems cover all three lines of
sight, suggesting that the gaseous structures giving rise to MgII
complexes are larger than $\sim$ 0.5 kpc. However, in most cases it is
difficult to trace {\it individual} MgII `cloudlets' over distances
larger than 200-300 $h_{50}^{-1}$ pc, indicating that typical sizes of
the MgII cloudlets are smaller than the sizes inferred earlier for the
individual clouds of high ionization gas seen in CIV absorption.  We
tentatively interpret the absorption pattern of the strongest MgII
system in terms of an expanding bubble or galactic wind and show that
the possible loci occupied by the model bubble in radius-velocity space
overlap with the observed characteristics of Galactic supershells.

\end{abstract}


\newpage

\clearpage

\section{Introduction}

Absorption lines in the spectra of the separate images of
gravitationally lensed QSOs provide a unique opportunity to investigate
the detailed structure of the interstellar gas and the intergalactic
medium on scales of ten parsecs to a few tens of kpc. The ultimate goal
of such observations is to study the evolution of the microstructure of
the gas in galaxies and in the intergalactic medium from the epoch of
formation of the first luminous objects in the universe down to the
present epoch. For the past few years we have undertaken a survey of
lensed QSOs with the HIRES spectrograph (Vogt et al 1994) on the Keck I
telescope on Mauna Kea.  The first paper in the present series (Rauch,
Sargent and Barlow 1999) analyzed a low ionization absorption system at
$z_{abs} = 3.538$ on a scale of only 26h$^{-1}_{50}$ parsecs. The
second paper (Rauch, Sargent and Barlow 2001) investigated the
structure of C IV absorbing clouds, including an estimate of the rate
of input of mechanical energy, while the third paper (Rauch, Sargent,
Barlow and Carswell 2001) set a limit on inhomogeneities in
the ``Lyman alpha forest'' clouds at redshift $z \sim 3.3$.

In the present paper we turn our attention to low ionization systems
which are characterized by MgII or FeII absorption lines.  This gas is
likely to be denser, more metal-rich,  and more closely associated with
galaxies than either the average Lyman $\alpha$ forest or the high
ionization CIV absorption systems.

At intermediate redshifts MgII absorbers have usually been interpreted
as the large gaseous halos of luminous galaxies. The sizes of entire MgII
complexes are relatively well constrained:  Smette et al (1995), from
observations of HE1104-1805A,B, estimate lower limits of $>20
h_{50}^{-1}$ kpc for the diameters of MgII systems (see also the high
resolution study of the z=1.66 system in that QSO by Lopez et al
1999).  Monier et al (1998) find an upper limit of $20h_{50}^{-1}$ kpc
for a strong Lyman limit absorber. HIRES spectra we have obtained of
the two lines of sight to UM673 A and B (to be published) show three
strong MgII absorption systems. The one at z=0.426 (transverse
separation between the lines of sight: $14.6 h_{50}^{-1}$ kpc) shows
absorption in only one of the images, whereas in the other two systems
(z=0.492, sep.  = $ 15.5h_{50}^{-1}$ kpc; z=0.564, sep. = $
13.9h_{50}^{-1}$ kpc) there is MgII absorption of similar strength in
both sight lines.  Surveys for galaxies in the vicinity of MgII
absorption systems and attempts to reconcile the total absorption cross
section with the luminosity function of galaxies (Bergeron \& Boiss\'e
1991; Steidel 1995; Bowen, Blades \& Pettini 1995; Churchill et al
1999) have shown that galaxies appear to be surrounded by halos of MgII
gas with radii estimated to be on the order of 60-130 $h_{50}^{-1}$
kpc.

The spatial extent of {\it individual MgII absorption components} as
opposed to entire MgII systems is somewhat harder to pin down.  The low
ionization system discussed by Rauch, Sargent and Barlow (1999) showed
that SiII and CII gas cloudlets (at least at z=3.6) can be as small as
a few tens of parsecs.  Petitjean et al (2000) have investigated
intermediate redshift MgII systems in a (spatially unresolved) HIRES
spectrum of the lensed BAL APM08279+5255 and conclude from the residual
intensity in the MgII doublets that the sizes of individual clouds are
less than $1.5 h_{50}^{-1}$ kpc.  The only other independent size
estimate on individual cloudlets comes from ionization calculations
(e.g., Bergeron \& Stasinska 1986; Steidel 1990), which, however, 
provide more indirect constraints.  Rigby et al (2002) using such
arguments, find that a subset of iron-rich MgII systems may be
consistent with cloudlets as small as 10 pc in diameter.

In the present paper we address the question of the small scale structure
of low and intermediate ionization gas in more detail.
In particular, we investigate the  absorption lines in the spectrum of
the "Einstein Cross" in which Q2237+0305 ($z_{em} = 1.69$) is
gravitationally lensed by a 15th magnitude galaxy with a redshift z =
0.0390 (Huchra {\it et al.} 1985). Subsequent work (Yee 1988) revealed
that the galaxy is a barred Sb with a ring, and has a disk scale length $r_0 = 9.4"$
and an effective radius for the bulge $r_e = 1.9"$. 
Q2237+0305 turned out to have four images roughly arranged in the form of a cross
centered on the nucleus of the galaxy (Yee 1988; see his fig.4). The typical
separation between images is less than 2". Crane et al. (1991) obtained B =
17.96 magn. for component A, B = 17.82 magn. for B, 18.66 magn. for C
and 18.98 magn. for image D.  Later observations showed that the
relative brightnesses of the images is subject to changes, in part due
to microlensing (Irwin et al 1989).  Note that the flux ratios obtained
during the current observation (in the caption of Fig. 1) are very different
indeed.

 The original discovery spectra only showed self absorption in
the Lyman alpha and C IV emission lines.  Hintzen {\it et al.} (1990)
searched for but failed to find Ca II H and K absorption due to the
lensing galaxy in the spectra of the quasar images.  However, they
detected Mg II and Fe II absorption at $z_{abs} = 0.5664$ and a C IV
complex at  $z_{abs} = 1.694$ and 1.697.

In section 2 we describe the high resolution observations of
the three brighter components of Q2237+0305 with HIRES.  
Section 3 presents the detection of CaII absorption from the Milky Way and Section 4
describes a CaII system arising in the lensing
galaxy. Section 5 discusses the properties of a peculiar, strong MgII
absorption and tries to give a tentative interpretation. 
Sections 6 and 7 briefly describe the other two low ionization systems
towards Q2237+0305, and section 8 gives the conclusions.

\section{Observations and data analysis}

Q2237+0305 has four images roughly arranged in the form of a cross
centered on the nucleus of a 15th magnitude galaxy. A picture of the
lensed QSO taken off the Keck HIRES guider is shown in Fig. 1 where
the four images are identified. Spectra of the three brighter images A,
B and C were obtained with HIRES in October 1998. The blue cross
disperser was used, giving a wavelength coverage of 3645-5204 \AA\ which
puts the C III] 1909 emission line at the red end of the range
covered.  We used a 0.86" slit, giving a resolution of 6.6 km
s$^{-1}$.  The exposures, broken into 3000 second sections, totaled
18,000 seconds on image A, 18,000 s on B and 15,000 s on C. 
A heliocentric correction was applied. All wavelengths quoted here
are vacuum values. The individual spectra of
each image were combined and scaled to the same continuum level in
order to compare the details of the absorption lines.  The absorption
systems were fitted with the Voigt profile fitting routine VPFIT
(Carswell et al 1991, 1992). Fitting components were put in until a
statistically satisfying $\chi^2$ value was achieved. Usually this
involved obtaining a reduced $\chi^2$ with a probability $> 1\%$ to have
ocurred by chance. The continuum level was treated as an additional
free parameter (see paper III) to reduce systematic errors that could
have crept in with the interactive continuum fitting.

The interpretation of this dataset is somewhat more challenging than
usual because the QSO images closely surround (to within an arcsecond)
the peaked emission from the bulge of the foreground lensing galaxy, so
there is a high background level which rapidly changes in the radial
direction (Fig. 1).  The data show several instances of absorption
lines which to the eye appear flat-bottomed and saturated while not
dropping down to zero intensity in the line center. This phenomenon is
most obvious in the strong absorption of the MgII complex at z=0.5656,
especially in images $B$ and $C$ (see Figs. 5 and 6 and below, and the discussion
in section 4.1), where there
appear to be partly flat-bottomed MgII 2796 lines with a residual
intensity of about 11\% (in image $B$). At values of 1.35 and 1.16 the
MgII doublet ratios for the strongest components in the $A$ and $C$ images,
respectively, also suggest partial saturation.  There are several
possible explanations for this behaviour: (a) the absorbing gas cloud
does not cover the continuum source of the background QSO and the
troughs are partly filled in with light by-passing the absorber; (b)
the absorption lines are {\it very} narrow and unresolved by the
instrument; and (c) there is a foreground light source which partly
fills in the absorption lines of systems arising between the foreground
source and the background QSO.  Possibility (a) is quite unlikely in
that all of the low ionization systems are very far away from the QSO
redshift. Case (b), unresolved narrow lines, may indeed by present,
but the flat bottoms of the absorption lines and the fact that
different lines often reach similar levels argues against it being the
main effect:  it would require a conspiracy between individual optical
depths and velocity separation between the components to produce such a
pattern. Possibility (c), pollution by the light of the bright
lensing galaxy  appears most likely:  inspection of various strong lines
in all three lines of sight shows that the residual intensities at the
line bottoms correlate with the local flux of the QSO (they are weaker
where the QSO is brighter, e.g., in the CIV emission line), as one
would expect if the contribution of the QSO continuum light to the
total spectrum increases relative to the light of the foreground
galaxy.  The relative contribution of galaxy and
QSO continuum to the flux at a given wavelength, $F_{gal}/F^{cont}_{QSO}$, can be
written in terms of the residual intensity at the bottom of an
apparently saturated absorption line, $F^{res}(\lambda)$, as
\begin{eqnarray}
\frac{F_{gal}(\lambda)}{F^{cont}_{QSO}(\lambda)} = \frac{F^{res}(\lambda)}{1 - F^{res}(\lambda)}
\end{eqnarray} 
For image $A$ the MgII 2796 line at z=0.5656 gives approximately
$F_{gal}/F^{cont}_{QSO} \sim 0.08$, while the strongest CIV 1548 line
at z=1.693 gives 0.04. The CIV z=1.693 line is an associated system
riding almost on top of the CIV emission line and hence may be a candidate
for partial continuum coverage. However, we ignore this effect here
since it can only make the line less black.  For image
$B$ the MgII 2796 at z=0.5656  gives approximately
$F_{gal}/F^{cont}_{QSO} \sim 0.11$, while the CIV 1548 line at z=1.693
gives again about 0.04.  In other words the contribution of the light
from the lensing galaxy to the integrated spectrum is up to about 11\%
near the MgII 2796 z=0.5656 system.

A naive measurement of the column densities without correcting for the
light pollution by the lens would underestimate the column density for
weak lines on the linear part of the curve of growth by on the order of
0 - 12\% (see section 4 below), but for saturated lines they may be
highly underestimated.  Therefore, the column densities and measurement
uncertainties in tables 1,2, 4 and 5 below (which reflect only the profile fits
without correction for the residual flux from the lens) should be
taken as lower limits. The exception is the $z=0.5656$ low ionization system
(section 4) where many lines are obviously saturated, so a correction was
attempted by simultaneously adjusting the zero level of the
spectroscopic flux while fitting the absorption line profiles. The
correction was possible because there were enough saturated regions to
constrain the zero flux level, but it could not be applied meaningfully
to the other absorption systems which appear to be mostly optically
thin and do not have enough transitions available. Thus table 3
contains actual estimates of the absorption line properties.

\section{CaII absorption from the Milky Way}

The CaII absorption from our Galaxy can be modelled satisfactorily with
a single velocity component in all three images (see the fit parameters
in table 1). A pairwise comparison between any two of the three sightlines is shown in
Fig.  2. The column densities and Doppler parameters are identical to
within the errors, and the projected velocities along the line of sight
differ by less than 2.9 $\sigma$ between the images.  In other words,
there is no significant difference in the appearance of the Galactic CaII systems in
the three lines of sight.  The mean weighted heliocentric velocity of
the three sightlines is $cz = -13.9\pm 0.5$ kms$^{-1}$. The
direction of the Q2237+0305 sight lines in galactic coordinates
($l^{II}$ = 71.8, $b^{II}$ = -46.1 deg) subtends an angle of $\sim 63$
deg with the heliocentric velocity vector of the local bulk flow of
interstellar cloudlets within $\sim 30$ pc ($v_{\mathrm helio}$ =
$-28.1\pm 4.5$ kms$^{-1}$; Frisch, Grodnicki, \& Welty 2002) so the
projected velocity of that flow relative to the 2237+0305 line of sight
would be -12.9 kms$^{-1}$. This is well consistent with the idea that
the CaII observed here arises from the cluster of local interstellar
gas clouds within 30 pc. The maximum angle between the three lines of sight to
Q2237+0305 is about 1.8 arcsec.
Thus, if the Galactic CaII absorber towards 22327+0305 resides within
30pc, our observation means that there are no significant detectable
differences in that particular CaII cloudlet over transverse
separations smaller than 54 astronomical units.

\section{High velocity CaII clouds related to the lensing galaxy (z=0.039)}

\subsection{Observed properties}

We detect an absorption system at z=0.039, the redshift of the lensing
galaxy 2237+0305.  The system is visible in all three lines of sight
and consists of moderately narrow ($7.6<b<24.3$ kms$^{-1}$), presumably
nonstellar Ca II H and K absorption lines. A comparison of the Ca II H
and K profiles in each pair of the three images is shown in Fig. 3.
Each line of sight is shown separately in Fig.4.\footnote{Note that the
wavelengths of Ca II H and K quoted in Fig. 2,3 and 4 are the {\it vacuum}
values.} Any underlying, broad stellar Ca II absorption feature due to
the lensing galaxy has been removed during the background subtraction
and by drawing the continuum such as to eliminate large scale
features.  The velocity scale in Figs. 3 and 4 is relative to the
lensing galaxy redshift $z_{abs} = 11696(\pm 40)$ kms$^{-1}/c = 0.039014$ given
by Foltz {\it et al.} (1992).

Fig. 3 shows that image $A$ has a prominent absorption feature at
+103 km s$^{-1}$ and a weaker feature at -91 km s$^{-1}$.  \footnote{To
avoid confusion we do not show the fits on top of the spectra but give
the fitting parameters redshift, Doppler parameter $b$ and logarithmic
column density $\log N$ and their $1-\sigma$ errors in table 2.} Image
$B$ has three (and possibly more) weak components at -288, -202, and
+105 kms$^{-1}$.  Image $C$ has two strong components at -153 and -95
kms$^{-1}$ and two additional weaker ones at +38 and +90 kms$^{-1}$.
Only the two strongest lines, the +103 kms$^{-1}$ component in image
$A$ and the -95 kms$^{-1}$ component in $C$ are strong enough to be
clearly present in both transitions of the CaII doublet, and it is
only  for those that we can be sure of them being CaII. Any of the
other lines could in principle individually be an unidentified
interloper from other transitions (there is one such case, marked by
an $x$ in the CaII 3970 regions in Fig. 3 which is the 1550\AA\ line of
a stray CIV system at z=1.661), but the line density is much too low
for this to explain the majority of components.

Using intermediate resolution spectroscopy of the integrated light from
the four QSO images, Hintzen et al (1990) have searched for CaII from
the lensing galaxy, deriving a $3-\sigma$ upper limit of 72 $m\AA\ $ on
the rest frame equivalent width of the CaII K line. They compared this
limit with the equivalent width of CaII from the thick disk in our
Galaxy and concluded that the absence of strong CaII rules out the
presence of a thick disk near the center of 2237+0305.  Calculating the
CaII 3935 \AA\ equivalent widths from the column densities given in
table 2 we get $W$ = 97$\pm 4$, 122$\pm 12$, and 424$\pm 9$ $m\AA $,
for the total equivalent widths of all the fitted components in images
$A$, $B$, and $C$, respectively. The unweighted mean of the three
images is 214 $m\AA $.  These values are not corrected for the zero
level problem mentioned above but if anything the real values can only
be larger (probably by not more than 12\%).  The equivalent widths are formally inconsistent with the
null detection of Hintzen et al.  However, their result is an average
of the CaII absorption along all four lines of sight which they
observed through a 2.5" diameter circular aperture, whereas our
observation was done only for the brightest three images individually
with a slit 0.86" wide.  The fourth line of sight (included in their
average but not in ours) may not have any absorption, which would
reduce the discrepancy.  Another possible explanation has to do with
the uncertainties due to pollution of the QSO spectra by the bright
foreground galaxy.  Hintzen et al used an aperture larger by more than
a factor 8 than ours (assuming that we had 0.7" seeing), and the depth
of the stellar H and K lines in their spectrum shows that the galactic
contribution to the QSO spectrum was severe. Hintzen et al dealt with
this problem by subtracting sky "from a position 36 arcsec offset from
the QSO", and "subtracting a suitably scaled spectrum of the M31
globular cluster K225-B280" to remove the broad stellar H and K lines.
In our case, the sky background (including the stellar contribution)
was subtracted in windows immediately adjacent to the QSO image. We can only
speculate that given our own problems  in subtracting the spatially
fluctuating sky background (discussed in section 2) it is conceivable
that Hintzen et al may have had similar ones. They may have 
subtracted less of the stellar background than
we did which would have led to them underestimating the depth of the
absorption and thus the CaII equivalent width.

In any case, CaII is detected and it appears that the measured total
equivalent width is not dramatically different from what is seen in the
Galaxy: Bowen (1991) found the typical high-latitude CaII equivalent
width distribution to peak at $W \sin b = 110 m\AA\ $
for lines of sight going outward from the Galactic disk. In the present
case the lensing
galaxy at z=0.039 is inclined by 60 degrees (Irwin et al 1989) with
respect to the plane of the sky so the equivalent width expected
based on the Bowen study would be $W = 220 m\AA\ $.  Our 
measurement gives the two-sided equivalent width (intersecting the
whole galaxy) and translates into 214/2 = 107 $m\AA\ $ for a line of
sight going out from the central plane of the galaxy.  This is about
half of what is seen in the Milky Way for the thick disk.
However, most of the absorption components in the HIRES data
formally qualify as {\it intermediate} or {\it high velocity clouds}
(e.g. Wakker 2001), so a comparison of their equivalent width with a subset
of higher velocity clouds in our Galaxy would be more appropriate.  Galactic
intermediate velocity clouds studied by Morton \& Blades (1986) lead to
an average of $100\pm 50$  $m\AA\ $ (Hintzen et al. 1990), similar to the
present result, 107 $m\AA\ $.

We do not have information on the HI column densities so we cannot
obtain the gas phase abundances for these clouds.  If the abundances
measured in the interstellar medium of the Galaxy were similar to those
in the 2237+0305 galaxy, then one can use the CaII-HI abundance
relation of Wakker \& Mathis (2000) to determine the HI column
densities.  According to their formula, the CaII column densities in
table 2 would roughly correspond to logarithmic HI column densities
between 18.6 and 20.1, with the exception of the strongest component in
the $C$ image (at -95 kms$^{-1}$). This one formally gives a  very high
HI column (23.1) but it is beyond the range of validity of the
Wakker-Mathis relation, and may be more indicative of a higher than usual
CaII gas phase abundance.

\subsection{Origin of the CaII velocity structure}

The velocity structure of the CaII system
is different from that expected from absorption arising in a thick
disk: 

The center of CaII absorption for all CaII components in all three
sightlines (defined as the column density weighted mean velocity
relative to the redshift of the lensing galaxy) occurs at -71
kms$^{-1}$ with respect to the Foltz et al (1992) velocity of
recession, $11696 (\pm 40)$ kms$^{-1}$, or at +4 kms$^{-1}$ relative to
the HI 21cm velocity, $11621 (\pm 40)$ kms$^{-1}$ given by Barnes et al
(1999). Thus, within a 2-sigma deviation the column density weighted
velocity centroid of the CaII absorption agrees with the centroid of
the stellar and HI mass distribution.  However, CaII absorption from
the disk of our Galaxy seen against high latitude stars tends to have a
strong component within about 10-20 kms$^{-1}$ of the local standard of
rest (e.g., Greenstein 1968; Blades \& Morton 1983; Songaila et al
1986; D'Odorico et al 1989; Robertson et al 1991; Meyer \& Roth 1991;
Ho \& Filippenko 1995; this paper). Similar CaII absorption near
systemic galactic velocities is also seen towards other galaxies, e.g.,
against SN1989M, SN1993J, and SN1994I in the disks of M81 (Vladilo et
al. 1994), NGC4579 (Steidel, Rich \& McCarthy 1990), and NGC5194 (Ho \&
Filippenko 1995), respectively, and towards the nucleus of M31 (Morton
\& Andereck 1976).  This is in contrast to the present case where there
does not seem to be a zero velocity component in any of the three
sightlines, no matter which of the above velocity centroids for the
galaxy is adopted.

Furthermore, the spread in velocity of the CaII components is larger
than expected if caused by the disk.  The total unweighted velocity widths
(highest minus lowest velocity component) are $\Delta v_A = 193$,
$\Delta v_B=393$, and $\Delta v_C = 243$ kms$^{-1}$ for the three
images $A$, $B$, and $C$, respectively.  Even if only the relatively
certain strong components at -153 kms$^{-1}$ (C image) and +103
kms$^{-1}$ (A image) are used the velocity difference is at least 256
kms$^{-1}$.   The typical line of sight from the source QSO has an
impact parameter of $\sim 1"$ or 1.1h$^{-1}_{50}$ kpc.  Therefore,  the
three lines of sight pass through the bulge of the lensing galaxy,
rather than the disk, and it is clear that the rotation of the disk
cannot be responsible for the large velocity widths even if it extended
to the very center of the galaxy.  Thus we agree with Hintzen et al,
albeit for different reasons, that this case does not look like thick
disk absorption.

The origin of the large velocity width of the CaII absorption is not
obvious.  The width of the absorption system is not unlike what is seen
in sightlines towards the Magellanic clouds or in the Magellanic stream
(Songaila 1981; West et al. 1985;  Songaila et al 1986).  It is
possible that a similar configuration involving tidal gas flows (see also
Bowen et al. 1994)
exists here as well.  In their search for HI emission in the vicinity
of the lensing galaxy, Barnes et al (1999) have detected a group of
galaxies which contains the lens. The two objects nearest to the lens (in
projection), the dwarf irregulars NW1 and NW2 (in their nomenclature),
are separated from it by about 150 kpc and 125 kpc in the plane of the sky, and
in velocity of recession by -380 and -264 kms$^{-1}$, respectively. It
is conceivable that this group may produce a tidal feature like the
Magellanic stream, but the negative velocity differences of the closest
objects with respect to the lensing galaxy 2237+0305 may be
inconsistent with the positive velocities of some of the CaII
components. More suitable but fainter satellites of the 2237+0305 galaxy may have
eluded detection.

Even with the lack of evidence for a tidal mechanism there still
remains a wealth of possible causes as have been discussed in the
general context of Galactic high velocity clouds (which we do not wish
to enumerate  here; for a review see Wakker, van Woerden \& Gibson
1999, and references therein).  One possibility not yet mentioned is
that the CaII clouds may arise directly in the bulge of the z=0.039
galaxy.  Barnes et al (1999) have argued that the stellar velocity
dispersion measured by Foltz et al (1992), corrected for observational
effects, agrees with various mass models in giving a line of sight
velocity dispersion $\approx  145$ kms$^{-1}$ for the bulge.  The
velocity differences between CaII components quoted above exceed this
value by at least 110 and possibly as much as 250 kms$^{-1}$ (if all
CaII components are real). To account for the additional velocity
dispersion one could envoke outflows (winds, SN bubbles) from the
bulge, or there could be other hydrodynamic effects in the bulge that
stir up the gas of an otherwise quiescent disk. Morton \& Andereck (1976) have observed
additional velocity components at -260 and -450 kms$^{-1}$ against the
nucleus of M31 and argue that they are likely to be caused by
outflows. However, there is no evidence for nuclear activity in
the galaxy 2237+0305.

Finally there are the size constraints from the three lines of sight.
While the column density of the components  is rather different between
the sightlines there is a hint that at least two components of the C
image (-95 and +90 kms$^{-1}$ seem to correspond to weak features in
the A image (at -91 and +103 kms$^{-1}$).  Accordingly, this particular
system does have at least some coherence over 1.4 kpc, whereas the
average cloud size (if defined as the distance over which the column
density varies by 50\% ) may be much smaller. Coherence on kpc scales
may argue for clouds outside the bulge or disk, but the evidence is
weak in the present case.  Summarizing, the appearance of the z=0.039
system is consistent with that of high velocity clouds in the
Galaxy. The velocity structure of the CaII components and the apparent
hole in any CaII absorption by a thick disk argues for a kinematic
connection of the CaII cloudlets with the bulge and / or locations
above the plane of the lensing galaxy.

\section{The MgII-FeII system at $z=0.5656$: expanding shell, collapsing sheet ?}

\subsection{Observational properties}

This system is the strongest absorption complex in the spectral region
covered by our observations (Figs. 5 and 6). The appearance of the MgII
2796, MgI 2853, and FeII 2600 absorption lines makes it likely that the
structure is a strong Lyman limit or damped Ly$\alpha$ system with the
hydrogen predominantly neutral:  first, MgI is rather strong (e.g.,
$\log N$(MgI) - $\log N$(MgII) = $-1.42\pm 0.08$ for the reddest
MgI/MgII component in the A image (z=0.5669)).  Photoionization models
with CLOUDY (assuming an $\alpha=-1$ powerlaw ionizing background with
intensity $J(912)=10^{-21}$ at $z=2.5$, blueshifted to z=0.56) show that
for a strong Lyman limit system (N(HI)=$10^{18}$ cm$^{-2}$) the total
gas density would have to be rather high ($\geq 350 $ cm$^{-3}$) to get
a MgI/MgII ratio as high as observed.  For a damped Lyman $\alpha$
system with N(HI)=$10^{20}$ cm$^{-3}$, the density would need to be
only as high as $0.5$ cm$^{-3}$.  Second, strong FeII and MgII lines
often indicate damped systems (Boiss\'e et al 1998; Rao \& Turnshek
2000).The total equivalent widths of MgII 2796 and FeII 2600 for the
z=0.5656 system in line of sight $A$ are 1.20 \AA\ and 0.73
\AA\ respectively (after correction for the zero level).  According to
Rao \& Turnshek, 50\% of those systems with EW(MgII 2796) $\geq 0.5
\AA\ $ and with EW(FeII 2600) $\geq 0.5 \AA\ $ are damped (N(HI)$\geq
2\times 10^{20}$ cm$^{-2}$).

The $z=0.5656$ absorption system shows an interesting degree of
coherence between the three lines of sight (Figs. 5,6):  in MgII 2796, two
massive groups of components (near 130 and 175 kms$^{-1}$ in the $A$  spectrum
(bottom)), separated by a gap in absorption near 150 kms$^{-1}$, can be
traced in all  three images (with the $A$ spectrum having an additional
strong component near 255 kms$^{-1}$). While the gap between the
components does not shift by more than a few kms$^{-1}$ between the
spectra the two components seem to get further apart in velocity space
as one goes from image $A$ to $C$ to $B$, i.e., in the order of
increasing distance from $A$. The velocity differences $v_A$, $v_C$,
and $v_B$, between the components are $v_A = 69.8$ kms$^{-1}$, $v_C =
81.4$ kms$^{-1}$, and $v_B = 102.5$ kms$^{-1}$).  The two massive
components are clearly composed of multiple components indicative of
differential motion in the gas; in particular the B spectrum shows a
pronounced ragged appearance.  The projected relative positions of the
images $A$, $B$, and $C$, can be expressed in terms of distances
$\overline{AB}$ (=0.66 kpc) and $\overline{BC}$ (=0.50
kpc)  between images $A$ and $B$,and $B$ and $C$, respectively, and
the angle $\angle(ABC) (=47.5\deg)$. 

The fits to the absorption line profiles are listed in table 3.
Experience shows that MgII, FeII and sometimes MgI have a very similar
component structure in velocity space even if the relative column
density levels differ.  Thus for each fitting component the redshifts
of MgII and FeII (and in the case of image $A$ also MgI) were tied
together, i.e. they were forced to have the same redshift. The redshift
errors are quoted in the entry for the MgII ion in table 3. Where the
corresponding other ions at the same redshift have nominal errors of
zero it should be understood that the MgII redshift error is actually
the combined error of the tied ions.

To correct for residual flux from the lensing galaxy the  zero level
within the fitting regions was adjusted as an independent variable
simultaneously with the profile fits.  The required adjustments of the
zero level relative to a unit continuum in the three regions containing
the MgII doublet and MgI (region 1), the FeII 2383 line (region 2), and
the FeII 2600 line (region 3), turned out to be $7.9\pm 1.2$\%,
$-0.02\pm 4.8$\%, and $-0.07\pm 7.9$\%, respectively, in the $A$
spectrum, $10.6\pm 1.4$\%, $12.2\pm4.8 $\%, and $5.5\pm 6.6$\% in the
$B$ image, and $11.6\pm 0.7$\%, $9.5\pm 1.4$\% and $10.7\pm 1.4$\% in
the $C$ image. These values can be used as $F^{res}(\lambda)$ in eqn. 1
to obtain the ratio between galaxy and QSO fluxes which made it through
the spectrograph slit.

Limits on the gas temperature can be estimated from the line widths of
the MgII doublet and the FeII 2600 \AA\ line.  An upper limit on the
thermal width is given by the total line widths of the MgII lines in
Fig. 6. Limiting ourselves to the B image (which appears to have the
weakest lines, least likely to be saturated) a simultaneous fit to the
MgII doublet gives a weighted mean Doppler parameter $5.3\pm 0.4$
kms$^{-1}$ for the components in the $z=0.5656$ complex, corresponding
to an upper limit on the  temperature $T< 4.1\times 10^4$ K. A more
stringent limit may be obtained from comparison between two ions with
different atomic masses, MgII and FeII.  For example, in Figs. 5 and 6
the MgII 2796 spectrum of the $B$ image shows a group of 3 -- 4 ragged
components between 140 and 210 kms$^{-1}$.  The three strongest MgII
components are well matched by corresponding FeII 2600 \AA\ lines (see
also Fig.  5).  Another good case for an unblended line occurs near 250
kms$^{-1}$ in the $A$ image.  Using those 3 MgII/FeII components from
image $B$ where $b$(MgII)$\geq$ $b$(FeII) together with the abovementioned
line from the $A$ spectrum, we employ the method discussed by Rauch et
al (1996) to disentangle thermal and turbulent contributions to the
total Doppler width.  We have omitted two other cases where FeII was
broader than MgII as not providing information. The mean weighted
thermal MgII Doppler parameter turns out to be $4.0\pm 0.8$ kms$^{-1}$,
corresponding to a temperature $2.3\times 10^4$ K. Thus there appears
to be a substantial thermal contribution to the broadening (on the
order of a a few times $10^4$ K as opposed to a few hundred Kelvin or less
which would not show up observationally).  Unfortunately, the large
errors and the line widths being close to the resolution limit preclude
any further constraints.

\subsection{Modelling the absorbing structure as a moving shell of gas}

It is impossible to say with certainty how this particular absorption
system arises.  However, the double component structure mentioned above,
the increase of velocity separation between the two components with
transverse separation between the lines of sight, and the large 
coherence length (Fig. 6) are suggestive
of the three lines of sight intersecting two walls of a
coherent, expanding (or collapsing) "shell" (Fig. 12). Earlier we described a
small low ionization absorption structure in a high redshift galaxy
towards Q1422+231 which had several of the absorption characteristics
expected from an expanding shell (paper I).  The
size of that structure, however, was at least one to two orders of
magnitude smaller than the current object, which must be larger than
several hundred parsecs in order to cover all three lines of sight. It
is conceivable that this absorption system could be associated with
galactic HI shells and supershells which are known to be at least that
large (e.g., Heiles 1976,1979; Deul \& den Hartog 1990; Kennicutt et al
1994; Oey \& Clarke 1997). Such structures may be caused by SN
explosions or wind blown bubbles (Heiles 1979) or by the infall of
gas from the intergalactic medium e.g., in the form of high velocity
clouds (Oort 1967).  Recently, Bond et al (2001a,b), in a detailed
analysis, have given plausible arguments for at least some MgII
absorbers being caused by galactic ejecta, i.e., super-bubbles and
-winds. The line of sight component structure of the $z=0.5656$ system is quite
similar to that of several MgII systems in Bond et al. 2001b. 

Here we attempt to model the absorption system as a spherical bubble in
isotropic expansion (or contraction).  Having the velocity information
from three lines of sight we can infer the radius and expansion
velocity of the bubble from observations of the same system intersected
by multiple lines of sight.  The measured input parameters for such a
calculation are the three velocity differences between the main
components and the three transverse distances between the lines of
sight. Unfortunately, these constraints are not enough to determine
both radius $R$ and expansion velocity $v$ independently, and only the
relation $R(v)$ can be obtained. The line of sight to the fourth
($D$) image also would have to be observed, which was not attempted by
us because of the small separation and the faintness of image $D$. The
Appendix shows how to calculate the relation $R(v)$, subject to the
observational constraints. The result is given in the form of a diagram
in Fig. 11, and the geometry is shown in Fig. 12.  There is a broad minimum in the expansion velocity leading
to values below 100 kms$^{-1}$ for bubble radii between a few hundred
parsecs and a few kiloparsecs, with the minimum expansion velocity, $v
= 55$ kms$^{-1}$, occuring at a radius $R = 0.83$ kpc.  At very small
radii the diameter of the bubble approaches the transverse separation
between the lines of sight. Such small bubbles require not only an
increasingly unlikely alignment between the bubble walls and the lines
of sight to still cover all three of them. They also lead to a
near-grazing incidence of the lines of sight with respect to the bubble
and they require very large expansion velocities to reproduce the
velocity splitting seen between the components.  At the opposite end,
for very large bubbles with $R >> 1$ kpc, the expansion velocity also
would have to be very high: the spatial curvature of the bubble wall
between the lines of sight would be smaller with increasing radius, and
the differential velocity splitting between the major components seen
in Fig. 6  can only be reproduced with an ever increasing expansion
velocity.

Galactic and local extragalactic superbubbles tend to show maximum
sizes of about 1-2 kpcs, and expansion velocities are typically below
150 kms$^{-1}$ (Heiles 1979; Tenorio-Tagle \& Bodenheimer 1988; Martin
1998).  Thus the parameter space of observed superbubbles shows
considerable overlap with the allowed combinations of $R$ and $v$ for
our model (hatched area in Fig.  11).  
The observed total MgII column densities in the two main components are
$2.7\times 10^{13}$ and $3.3\times 10^{13}$ cm$^{-2}$
for the blue and red component in the $A$ image, $1.8\times 10^{14}$ and $2.7\times 10^{13}$ cm$^{-2}$ for $B$, and $7.8\times10^{14}$ and $4.5\times 10^{13}$ cm$^{-2}$ for the  $C$ image,
respectively. We can obtain a crude upper limit on the size of the bubble
by assuming that the absorption is arising entirely in a bubble shell composed of gas swept up by
a wind outside the disk's interstellar medium (assuming subsequent cooling to produce MgII as the dominant Mg ionization state). Then
to produce a MgII column density as strong as observed the gas would have to come from a volume with radius
\begin{eqnarray}
R = 3 \frac{N}{n}= 2\ {\mathrm kpc} \left(\frac{N_{MgII}}{10^{14}{\mathrm cm}^{-2}}\right)\left(\frac{n_{tot}}{10^{-2}{\mathrm cm}^{-3}}\ \ \frac{A_{Mg}}{3.3 \times10^{-5}}\ \ \frac{Z/Z_{\odot}}{0.3} \ \ \frac{f_{MgII}}{0.5}\right)^{-1}.
\end{eqnarray}
Here $N$ is the total column density of the shell  and $n$ is the
background number density of the gas to be swept up, where we assume
the value for the halo gas density from model A1 by Suchkov et al.
(1994). $N_{MgII}$ is the column density,  $A_{Mg}$ the relative
elemental abundance of Mg, $Z/Z_{\odot}$ the metallicity in terms of
solar abundances, and $f_{MgII}$ the ionization fraction of MgII. The
latter is quite uncertain.  $Z/Z_{\odot}$ is probably between 0.1 and 1,
judging from the abundances of damped Ly$\alpha$ systems at low z
(e.g. Boiss\'e et al 1998). The radius $R$ is again fully consistent
with observed superbubbles (if much of the interstellar medium is swept up
as well the radius could be smaller, of course).

It is conceivable though that the $z=0.5656$ system is not produced in the
bubble walls of a smallish ($\sim 1$ kpc) shell but rather by
superwinds. Then the absorption would come from the dense shreds and
filaments formed when the interstellar medium gets entrained by the hot
outflowing wind (cf Bond et al 2001; Suchkov et al 1996; Heckman et al
2000, 2001; Mori, Ferrara \& Madau 2002).  The ragged component
structure seen here may be caused by differential motion between 
gas clouds accelerated by the surrounding wind.

The possibility of the absorber being caused by (symmetric ?) infall of
gas  has also been discussed, e.g., in the form of high velocity clouds
which may be sweeping up galactic material.  Apparently sizes and
infall velocities of known high velocity clouds are sufficient to
produce large scale features in the galactic gas upon impact and may
even be able to more easily satisfy the energetic requirements of the
largest shells than collective supernova explosions (e.g., Oort 1967;
Tenorio-Tagle 1981; Tenorio-Tagle \& Bodenheimer 1988 and references
therein; Rand \& Stone 1996).  At high redshift low ionization metal
absorption systems from infalling gas are expected to arise as a
natural by-product of galaxy formation (Rauch, Haehnelt \& Steinmetz
1997).  Once the gas makes it past the accretion shock surrounding the
forming galaxies it cools and can be seen in MgII or damped Ly$\alpha$
absorption.  However, it is not clear whether cosmological infall of
pre-enriched material can reproduce all the small scale density and
velocity structure seen in real metal absorption systems (paper I,
II).

On the basis of the observations presented here and the
rather idealized theoretical models we can currently not elucidate the
origin of the z=0.5656 system any further.  It would be interesting to
study more MgII systems in the spectra of lensed QSOs to see whether
there is a statistical difference in spatial and velocity extent
between the two-component systems like the one seen here and randomly
selected systems.

\section{The z=0.82743 system}

This system also has a prominent double-component structure that appears in
all lines of sight, albeit shifted and with rather different column densities
(Figs.  7 and 8). The velocity separation $\Delta v$ between the two
most prominent MgII lines in each of the A, C and B images is 35.0, 47.9 and 18.6
kms$^{-1}$, respectively. The velocity structure is more complicated than
in the z=0.5656 system in that here are also shifts of the column-density
weighted velocity along the line of sight, $\overline{v}$, between the
lines of sight, which amount to $\overline{v_C} - \overline{v_A} =
-32.2$ kms$^{-1}$, $\overline{v_B} - \overline{v_A} = 16.5$
kms$^{-1}$, and $\overline{v_B} - \overline{v_C} = 64.2$
kms$^{-1}$.

\section{The z=0.97163 system}

For this system (Figs. 9 \& 10) only FeII lines are covered. Again there is some resemblance
between the absorption patterns for the three lines of sight. 
The differences between the column density weighted velocities along the line
of sight are $\overline{v_C} - \overline{v_A} =
4.1$ kms$^{-1}$, $\overline{v_B} - \overline{v_A} = -13.4$
kms$^{-1}$, and $\overline{v_B} - \overline{v_C} = -17.4$
kms$^{-1}$. The differential motion in this and the z=0.8274 system are small
enough to be attributed to rotation, .e.g., of an entire gaseous halo
(e.g., Lanzetta \& Bowen 1992; Steidel et al 2002), but even for the simplest
rotation model there are too many free parameters to be constrained by the anecdotal
evidence presented here, and a larger sample is clearly required.

\section{Conclusions}

(1) CaII absorption from the Galaxy is found in all three lines of sight to
Q2237+0305 with an average heliocentric velocity $-13.9\pm 0.5$
kms$^{-1}$ and without significant differences between the lines of sight.
This velocity is consistent with the absorption arising in the
local bulk flow in the ISM within $\sim 30$ pc from the sun (Frisch et al 2002).
In that case the largest angular separation between the sightlines
corresponds to $< 1.8" \times 30$ pc = 54 astronomical units. Apparently, the Galactic
CaII cloud seen toward Q2237+0305 is smooth on scales smaller than that.

(2) We have detected a high velocity cloud system related to the lensing
galaxy 2237+0305 (z=0.039). Unlike CaII systems in our Galaxy this one
does not show a strong component at the systemic velocity of the galaxy
as measured either from the stellar or HI centroid. The system arises
within impact parameters on the order of  1 kpc to the center of the
galaxy (i.e., behind and / or in front of its
bulge), so the large velocity differences among the components
along individual lines of sight and between different lines of sight cannot
be a result of the rotation of the galactic disk. 
The absorption line strengths are similar to those measured for intermediate
and high velocity clouds in our Galaxy. The system may have an origin
in the bulge or halo of the galaxy 2237+0305.

(3) Three MgII or FeII selected low ionization systems not associated with
the lensing galaxy show coherent
absorption over the transverse separation between the lines of sight,
ranging from about 200 - 600 $h_{50}^{-1}$ pc. This is a strict lower
limit to the size of the the gaseous structures giving rise to the
entire MgII system: they must be at least that large to cover all three
lines of sight. The closeness of the lines of sight in the present case
does not permit an upper limit to be obtained.

(4) individual MgII components are much smaller than entire MgII systems:
interpreting the separation where the optical depth between two lines
of sight differ by 50\% as a minimum `size' of the clouds, the three
systems examined here show fluctuations at that level for {\it
separations at least as small as $200-300 h_{50}^{-1}$ pc}. In fact,
several components do not show up at all in more than one image (Figs.
5-10).  This is consistent with the Petitjean et al upper limit but it
indicates that many of the cloudlets are actually much smaller than a
kpc. This result appears in contrast to size estimates for CIV
systems.  Individual CIV clouds appear to be larger than  about $300
h_{50}^{-1}$ pc (paper II).  Ionization calculations place the MgII gas
at a higher density than the more common, higher ionization, CIV gas
clouds.  Thus it is plausibly that the MgII cloudlets are smaller than
those clouds as well.

(5) We have tentatively interpreted the low ionization system at $z=0.5656$
as a moving shell, possibly caused by an expanding superbubble or galactic wind.
The large scale coherence of the system and the differential 
velocities between the absorption in the three lines of sight are consistent
with supershell parameters similar to those observed in the Galaxy and
in local dwarf galaxies. In this picture the two MgII components observed
may correspond to the wall of the cooling shell at the interface with the
intergalactic medium and / or the entrained material within the hot bubble.
Observations of transitions associated with higher ionization stages, especially
CIV and OVI, in the UV should be able to shed further light on the nature
of this system.

\acknowledgments
We thank Crystal Martin for helpful discussions, and Bob Carswell and
Gary Ferland for making VPFIT and CLOUDY, respectively, available.  At
Keck,  Gary Puniwai, Ron Quick, Gabrelle Saurage, David Sprayberry,
Wayne Wack, and Greg Wirth helped with the observations. The anonymous
referee is thanked for several helpful suggestions. MR is
grateful to the NSF for grant AST-0098492 and to NASA for grant
AR-90213.01-A.  The work of WLWS was supported by NSF grant
AST-9900733.

\pagebreak

\section{Appendix}

This is an outline  of the calculation of the relation between the expansion
velocity $v$ of a spherical shell, and its radius $R$, as a function of the observed velocity
spreads between absorption line pairs like the one in the $z=0.5656$ MgII system ($v_A = 69.8$ kms$^{-1}$, $v_B = 102.5$ kms$^{-1}$, $v_C = 81.4$ kms$^{-1}$) and the geometric constraints (the projected relative positions of the images $A$, $B$, and $C$,
as given by the distances $\overline{AB}$ (=0.66 kpc) and $\overline{BC}$ (=0.50 kpc), and the angle $\angle(ABC) (=47.5\deg)$).

There are four unknowns, which can be expressed as (for example)
$v$, $R$, $b$ (the impact parameter of line of sight
$B$ with respect to the center of the sphere), and $\phi$ (i.e., the angle between $b$  and $\overline{AB}$, the line connecting
lines of sight $A$ and $B$), and three equations connecting them (see below).
Thus we can only solve for a relation between
any two of these. The relation $v(R)$ is probably the most useful one as it determines
which values of size and expansion velocity are consisting with the observations.
The various geometric relations between these quantities can be written as:

\begin{eqnarray}
R^2 \left(1-\left(\frac{v_B}{2v}\right)^2\right) = b^2\ \ \ \ \ \ \ \ \ \ \ \ \ \ \ \ \ \ \ \ \ \ \ \ \ \ \ \ \ \ \ \ \ \ \ \ \ \ \ \ \ \ \ 
\end{eqnarray}
\begin{eqnarray}
R^2 \left(1-\left(\frac{v_A}{2v}\right)^2\right) = b^2 + \overline{AB}^2 + 2 b \overline{AB} \cos \phi \ \ \ \ \ \ \ \ \ \ \ \ \ \ \ \ 
\end{eqnarray}
\begin{eqnarray}
R^2 \left(1-\left(\frac{v_C}{2v}\right)^2\right) = b^2 + \overline{BC}^2 + 2 b \overline{BC} cos (\phi + \angle(ABC))
\end{eqnarray}

Elimination of the unwanted
parameters leads to the single equation
\begin{eqnarray}
R^{-4} (-8 \overline{AB}^3 \ \overline{BC}\ v^2 (4\ \overline{BC}^2 v^2 + R^2 (v_C^2-v_B^2))\ y + &&\nonumber \\
2\overline{AB}\ \overline{BC} R^2 (v_B^2 - v_A^2) (4\overline{BC}^2  v^2  +
R^2 (v_C ^2 -v_B^2)) y + &&\nonumber \\
16 \overline{AB}^4  \overline{BC}^2  v^4 + \overline{BC}^2 R^4 (v_B^2 - v_A^2)^2 + 
\overline{AB}^2 (16  \overline{BC}^4  v^4 + R^4   (v_B^2  
- v_C^2)^2 - \nonumber&& \\
8  \overline{BC}^2 R^2 v^2 (-v_C^2  - 
v_A^2 y^2 + 8 v^2 z^2 - v_A^2  z^2 + v_B^2 (1 + y^2 - z^2))))& = &0,
\end{eqnarray}
where $y = \cos(\angle(ABC))$, and $\ z = \sin(\angle(ABC))$.  
It can be solved for the velocity $v$. Of the four solutions, two are imaginary
and one describes a contracting ($v<0$) rather than an expanding sphere.
The remaining solution is well-behaved for $0.33$ kpc $< R < \infty$:
\begin{eqnarray}
v& =& 1/2\ [(\overline{AB}\ \overline{BC}^3 R^2 (v_A^2 - v_B^2) y + \overline{AB}^3 \overline{BC} R^2 (v_C^2 -v_B^2) y +  \nonumber \\
& &\overline{AB}^2 \overline{BC}^2 R^2 (v_B^2 (1 + y^2 - z^2) -v_C^2 - v_A^2) +\nonumber \\
& & [ ( \overline{AB}^2  \overline{BC}^2 R^4 ( - ( \overline{BC}^2 - 2   \overline{AB}\   \overline{BC}  y - 4  R^2  z^2 +  \overline{AB}^2) ( \overline{AB}^2 (v_B^2 - v_C^2)^2 -\nonumber\\
&& 2  \overline{AB}\  \overline{BC} (v_B^2 - v_A^2) (v_B^2 - v_C^2) y +  \overline{BC}^2 (v_B^2 - v_A^2)^2 ) + \nonumber \\
&&(\overline{BC}^2 (v_B^2 - v_A^2) y + 
\overline{AB}^2 (v_B^2 - v_C^2) y + \overline{AB}\  \overline{BC} (v_C^2 + v_B^2 (z^2 -1 - y^2 ) + v_A^2 ) )^2) ]^{1/2} ) / \nonumber \\
&&( \overline{AB}^2 \overline{BC}^2( \overline{BC}^2 - 2 \overline{AB}\ \overline{BC}  y - 4  R^2  z^2 + \overline{AB}^2))]^{1/2}  
\end{eqnarray}

Inserting the numerical values gives
\begin{eqnarray}
v = 0.5 \sqrt{4.878\times10^{15} \sqrt{\frac{R^4 (0.3959 + R^2)}{(9.3456 - 84.5541 R^2)^2}} +
\frac{3.4712\times10^{15} R^2}{(84.5541 R^2 - 9.3456)}},\label{vee}
\end{eqnarray}
where $R$ is now to be measured in units of kpc, and $v$ in cms$^{-1}$.
The relation $v(R)$ is shown in Fig. 11.

\pagebreak

\pagebreak

\clearpage

\clearpage

\begin{deluxetable}{ccccc}
\small
\tablewidth{0pt}
\tablenum{1}
\tablecaption{The Galactic CaII system}
\tablehead{
\colhead{LoS} &
\colhead{ion} &
\colhead{$z$} &
\colhead{$b$} &
\colhead{$\log N$} 
}
\startdata
A&CaII& -0.000043$\pm$ 0.000002& 15.21$\pm$ 1.06&   12.259$\pm$ 0.025\\
B&CaII& -0.000056$\pm$ 0.000004& 15.88$\pm$  1.73&  12.194$\pm$  0.039\\ 
C&CaII& -0.000048$\pm$ 0.000003& 18.01$\pm$  1.20&  12.256$\pm$ 0.024\\
\enddata
\end{deluxetable}

\clearpage

\begin{deluxetable}{ccccc}
\small
\tablewidth{0pt}
\tablenum{2}
\tablecaption{The z=0.039 (lens) system}
\tablehead{
\colhead{LoS} &
\colhead{ion} &
\colhead{$z$} &
\colhead{$b$} &
\colhead{$\log N$} 
}
\startdata
A&CaII& 0.038700$\pm$ 0.000006&  8.76$\pm$  2.76&  11.513$\pm$ 0.093\\
A&CaII& 0.039369$\pm$ 0.000002&  7.63$\pm$  1.00&  11.901$\pm$ 0.039\\
B&CaII& 0.038016$\pm$ 0.000014& 18.78$\pm$  7.12&  11.722$\pm$  0.175\\ 
B&CaII& 0.038316$\pm$ 0.000018& 24.34$\pm$  8.91&  11.753$\pm$  0.169\\ 
B&CaII& 0.039377$\pm$ 0.000013& 12.80$\pm$  6.05&  11.490$\pm$  0.171\\ 
C&CaII&  0.038483$\pm$  0.000006&   15.46$\pm$  2.84&  11.842$\pm$ 0.072\\
C&CaII&  0.038685$\pm$  0.000001&   10.18$\pm$  0.45&  12.513$\pm$ 0.016\\
C&CaII&  0.039146$\pm$  0.000010&   15.05$\pm$  4.63&  11.604$\pm$ 0.114\\
C&CaII&  0.039325$\pm$  0.000009&   17.06$\pm$  4.05&  11.753$\pm$ 0.093\\    \enddata
\end{deluxetable}

\clearpage

\begin{deluxetable}{ccccc}
\small
\tablewidth{0pt}
\tablenum{3}
\tablecaption{The z=0.566 system, with corrected zero levels}
\tablehead{
\colhead{LoS} &
\colhead{ion} &
\colhead{$z$} &
\colhead{$b$} &
\colhead{$\log N$} 
}
\startdata

A&  MgII&    0.565696$\pm$  0.000003&    4.16$\pm$     1.26&  11.801$\pm$ 0.059\\
A&  MgII&    0.566289$\pm$  0.000004&    8.92$\pm$     1.30&  13.020$\pm$ 0.105\\
A&  MgII&    0.566304$\pm$  0.000003&    3.79$\pm$     1.74&  12.844$\pm$ 0.196\\
A&  MgII&    0.566352$\pm$  0.000035&   22.91$\pm$     4.53&  12.987$\pm$ 0.044\\
A&  MgII&    0.566480$\pm$  0.000038&    5.94$\pm$     5.16&  12.383$\pm$ 0.829\\
A&  MgII&    0.566518$\pm$  0.000002&    4.72$\pm$     0.59&  13.484$\pm$ 0.189\\
A&  MgII&    0.566655$\pm$  0.000003&    9.49$\pm$     1.16&  12.303$\pm$ 0.039\\
A&  MgII&    0.566797$\pm$  0.000010&    8.32$\pm$     3.53&  11.622$\pm$ 0.119\\
A&  MgII&    0.566929$\pm$  0.000001&    4.97$\pm$     0.25&  12.730$\pm$ 0.022\\
A&  MgI &    0.566289$\pm$  0.000000&    8.80$\pm$     4.28&  11.329$\pm$ 0.195\\
A&  MgI &    0.566304$\pm$  0.000000&    0.40$\pm$     0.78&  12.026$\pm$ 2.909\\
A&  MgI &    0.566352$\pm$  0.000000&   27.50$\pm$    30.69&  10.945$\pm$ 1.180\\
A&  MgI &    0.566480$\pm$  0.000000&   24.91$\pm$    17.02&  11.296$\pm$ 0.425\\
A&  MgI &    0.566518$\pm$  0.000000&    5.01$\pm$     1.41&  11.459$\pm$ 0.097\\
A&  MgI &    0.566655$\pm$  0.000000&   65.20$\pm$   146.68&  10.929$\pm$ 1.058\\
A&  MgI &    0.566929$\pm$  0.000000&    7.23$\pm$     1.72&  11.311$\pm$ 0.080\\
A&  FeII&    0.565696$\pm$  0.000000&   32.19$\pm$    38.86&  11.974$\pm$ 0.829\\
A&  FeII&    0.566289$\pm$  0.000000&   11.14$\pm$     1.02&  13.365$\pm$ 0.078\\
A&  FeII&    0.566304$\pm$  0.000000&    1.72$\pm$     2.01&  12.751$\pm$ 0.239\\
A&  FeII&    0.566352$\pm$  0.000000&   26.25$\pm$     7.23&  12.991$\pm$ 0.192\\
A&  FeII&    0.566480$\pm$  0.000000&    8.54$\pm$     4.75&  12.764$\pm$ 0.607\\
A&  FeII&    0.566518$\pm$  0.000000&    4.76$\pm$     0.80&  13.405$\pm$ 0.154\\
A&  FeII&    0.566655$\pm$  0.000000&    8.30$\pm$     2.44&  12.313$\pm$ 0.160\\
A&  FeII&    0.566797$\pm$  0.000000&   21.01$\pm$    44.96&  11.891$\pm$ 0.802\\
A&  FeII&    0.566929$\pm$  0.000000&    4.80$\pm$     1.12&  12.496$\pm$ 0.082\\
\hline
B&MgII &    0.566180$\pm$  0.000002 &     4.37 $\pm$     0.58 & 14.222 $\pm$     0.472\\  
B &MgII  &  0.566270$\pm$  0.000003 &     8.88 $\pm$     0.71 & 13.100 $\pm$     0.034\\ 
B & MgII  & 0.566432$\pm$ 0.000003 &     4.67 $\pm$     0.85 & 12.311  $\pm$    0.057\\  
B&MgII &    0.566496$\pm$  0.000001 &     3.81 $\pm$     0.80 & 12.627 $\pm$     0.037\\  B&MgII &    0.566570$\pm$  0.000002 &     4.02 $\pm$     0.93 & 12.469 $\pm$     0.039\\ 
B &MgII &   0.566660$\pm$  0.000003 &     6.15 $\pm$     0.86 & 12.852 $\pm$     0.055\\
B &MgII &   0.566709$\pm$  0.000015 &     5.27 $\pm$     4.07 & 11.776 $\pm$     0.455\\
B& MgII &   0.566825$\pm$  0.000007 &     7.79 $\pm$     2.90 & 11.869 $\pm$     0.101\\
B&  MgII &  0.566940$\pm$  0.000012 &     7.83 $\pm$     4.24 & 11.658 $\pm$     0.146\\ 
B&MgII &    0.567536$\pm$  0.000000 &     7.05 $\pm$     4.79 & 11.392 $\pm$     0.193\\ 
B & MgII  & 0.568018$\pm$  0.000003 &     3.30 $\pm$     1.19 & 11.876 $\pm$     0.055\\  
B &MgI  &   0.566190$\pm$   0.000002&      6.62$\pm$      0.62&  11.906  $\pm$  0.033\\ 
B&MgI  &    0.566441$\pm$   0.000037&     46.67$\pm$     13.54&  11.759  $\pm$  0.120\\  
B & MgI &   0.566655$\pm$   0.000003&      1.73$\pm$      2.19&  11.266  $\pm$ 0.107\\  
B & MgI &   0.566860$\pm$   0.000017&     13.20$\pm$      5.31&  11.165  $\pm$ 0.154\\  
B & MgI &   0.567934$\pm$   0.000012&      4.59$\pm$      4.30&  10.816  $\pm$  0.196\\  
B & FeII &  0.566180$\pm$  0.000000 &     6.63 $\pm$     0.79 & 13.513  $\pm$  0.192 \\
B & FeII &  0.566270$\pm$  0.000000 &     8.52 $\pm$     1.07 & 13.308  $\pm$  0.076 \\ 
B & FeII &  0.566432$\pm$  0.000000 &    13.61 $\pm$     7.00 & 12.384  $\pm$  0.182 \\ 
B & FeII &  0.566496$\pm$  0.000000 &     2.36 $\pm$     0.92 & 12.780  $\pm$   0.139\\  
B & FeII &  0.566570$\pm$  0.000000 &     4.05 $\pm$     1.12 & 12.648  $\pm$   0.059\\ 
B & FeII &  0.566660$\pm$  0.000000 &     5.01 $\pm$     1.08 & 12.891  $\pm$  0.078 \\ 
B & FeII &  0.566709$\pm$  0.000000 &     1.48 $\pm$     6.81 & 11.801  $\pm$  0.404 \\ 
B & FeII &  0.566825$\pm$  0.000000 &    20.92 $\pm$     7.52 & 12.370  $\pm$  0.123 \\ 
B & FeII &  0.567536$\pm$  0.000000 &     2.69 $\pm$    10.75 & 14.098  $\pm$  13.197\\  

\hline
 
C &   MgII&     0.566236$\pm$    0.000001 &    5.13$\pm$ 0.58&   14.881$\pm$     0.423\\  
C &  MgII&      0.566283$\pm$    0.000004&    22.96 $\pm$ 8.74&   12.518$\pm$     0.404\\  
C &  MgII&      0.566331 $\pm$   0.000005  &   4.45 $\pm$ 2.58 &  12.297$\pm$     0.265\\  
C &   MgII&     0.566489$\pm$    0.000003 &   11.57$\pm$ 1.23&   13.137$\pm$     0.071\\ 
C &   MgII&     0.566555$\pm$    0.000002 &    9.93$\pm$ 2.02&   13.462$\pm$     0.098\\  
C &   MgII&     0.566607$\pm$   0.000003&     5.79 $\pm$ 3.28&   12.383$\pm$     0.531\\  
C &   MgII&     0.566871$\pm$    0.000005 &   17.30$\pm$ 6.48&   11.646$\pm$     0.137\\  
C &   MgI&      0.566247$\pm$     0.000003&    10.72$\pm$ 0.85&   11.776$\pm$     0.028\\  
C &   MgI&      0.566534$\pm$     0.000005&    18.37$\pm$ 1.65&   11.883$\pm$     0.033\\
C &   MgI&      0.566550$\pm$     0.000003&     1.26$\pm$ 1.95&   12.029$\pm$     1.738\\  
C &   MgI&      0.566895$\pm$     0.000019&    11.73$\pm$ 5.59&   10.956$\pm$     0.162\\  
C &   FeII&     0.566236$\pm$    0.000000 &    8.70$\pm$  0.32 &  13.812$\pm$     0.054\\ 
C &   FeII&     0.566283$\pm$    0.000000 &    1.21$\pm$  1.36 &  13.458$\pm$     1.268\\  
C &   FeII&     0.566331$\pm$    0.000000 &    5.56$\pm$  1.55 &  12.507$\pm$     0.089\\  
C &   FeII&     0.566489$\pm$    0.000000 &    8.78$\pm$  0.55 &  13.451$\pm$     0.034\\  
C &   FeII&     0.566555$\pm$    0.000000 &    1.77$\pm$  0.65 &  15.035$\pm$     1.194\\  
C &   FeII&     0.566607$\pm$    0.000000 &    4.76$\pm$  0.90 &  13.085$\pm$     0.045\\  
C &   FeII&     0.566871$\pm$    0.000000 &    0.18$\pm$  0.32 &  12.019$\pm$     0.727\\  
\enddata
\end{deluxetable}

\clearpage

\begin{deluxetable}{ccccc}
\small
\tablewidth{0pt}
\tablenum{4}
\tablecaption{The z=0.827 system}
\tablehead{
\colhead{LoS} &
\colhead{ion} &
\colhead{$z$} &
\colhead{$b$} &
\colhead{$\log N$} 
}
\startdata
A&MgII &  0.827437 $\pm$  0.000001& 2.53$\pm$0.18&  12.585$\pm$    0.039\\     
A&MgII &  0.827638 $\pm$  0.000004& 1.63$\pm$2.17&  11.520$\pm$    0.214\\
A&MgII &  0.827682 $\pm$  0.000060& 7.54$\pm$12.58&  11.112$\pm$    0.644\\
A&FeII &  0.827435 $\pm$  0.000002& 1.40$\pm$0.88&  12.160$\pm$    0.071\\
\hline
B&  MgII& 0.827646$\pm$   0.000004& 4.38$\pm$1.49&  11.636$\pm$    0.063\\
B&  MgII& 0.827797$\pm$   0.000001& 1.33$\pm$0.24&  12.434$\pm$    0.133\\
B&  MgII& 0.827910$\pm$   0.000001& 1.66$\pm$0.33&  12.315$\pm$    0.084\\
B&  FeII& 0.827795$\pm$   0.000003& 1.20$\pm$1.58&  12.089$\pm$    0.146\\
\hline
C&MgII& 0.827461$\pm$ 0.000008 &2.90 $\pm$3.27&  11.167$\pm$ 0.139\\ 
C&MgII& 0.827608$\pm$ 0.000004 &9.59 $\pm$1.14&  11.923$\pm$ 0.037\\ 
\enddata
\end{deluxetable}

\clearpage
\clearpage

\begin{deluxetable}{ccccc}
\small
\tablewidth{0pt}
\tablenum{5}
\tablecaption{The z=0.972 system}
\tablehead{
\colhead{LoS} &
\colhead{ion} &
\colhead{$z$} &
\colhead{$b$} &
\colhead{$\log N$} 
}
\startdata
A&FeII& 0.971635$\pm$   0.000001& 3.35$\pm$   0.23&  13.362$\pm$ 0.021\\  
A&FeII& 0.971668$\pm$   0.000010& 8.99$\pm$   0.94&  12.570$\pm$ 0.119\\
A&FeII& 0.971965$\pm$   0.000018& 4.06$\pm$   6.21&  11.266$\pm$ 0.265\\
A&FeII& 0.971866$\pm$   0.000020& 3.42$\pm$   8.19&  11.113$\pm$ 0.361\\
\hline
B&FeII& 0.971545$\pm$ 0.000002& 4.02 $\pm$  0.47&  12.505$\pm$  0.025 \\ 
B&FeII& 0.971646$\pm$ 0.000019& 6.57 $\pm$  5.81&  11.579$\pm$  0.211 \\
\hline
C&FeII& 0.971576$\pm$   0.000008& 4.95 $\pm$1.49& 12.153 $\pm$   0.246\\       
C&FeII& 0.971651$\pm$  0.000041& 9.44 $\pm$6.83&  12.028 $\pm$   0.333\\
C&FeII& 0.972003$\pm$  0.000007&  6.17 $\pm$1.96&  11.778 $\pm$   0.084\\

\enddata
\end{deluxetable}

\clearpage

\begin{figure}[t]
\figurenum{1}
\includegraphics*[scale=1.7,angle=-0.]{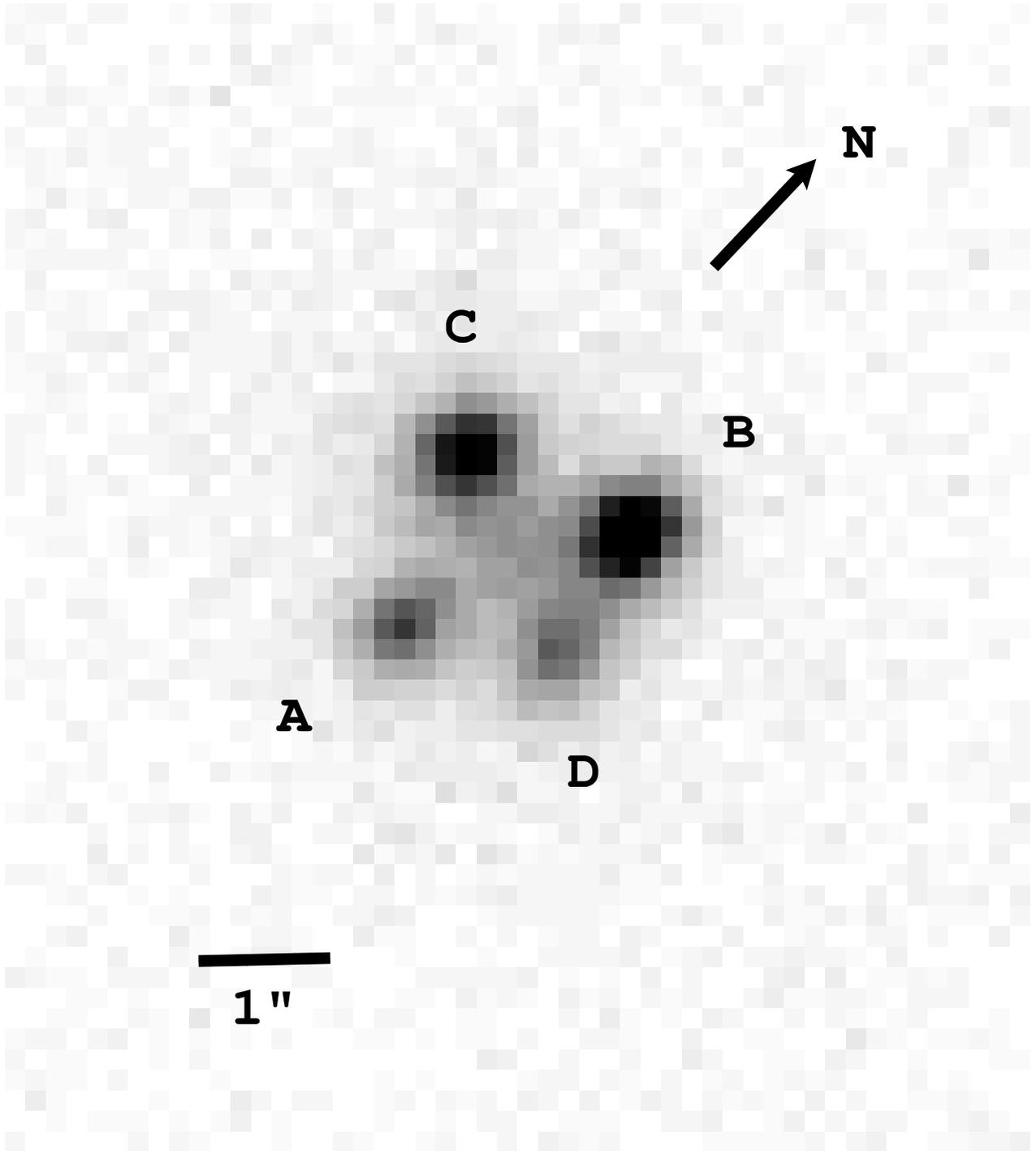}
\caption{\small  
Keck HIRES guider camera image of Q2237+0305. The exposure
time was 5 s, the picture was taken on October 21, 1998. The relative
flux ratios at that date were approximately $F_A : F_B : F_C : F_D = 1.0 : 2.1 : 1.7 : 0.7$.
}
\end{figure}

\begin{figure}[t]
\figurenum{2}
\includegraphics*[scale=0.8,angle=-0.]{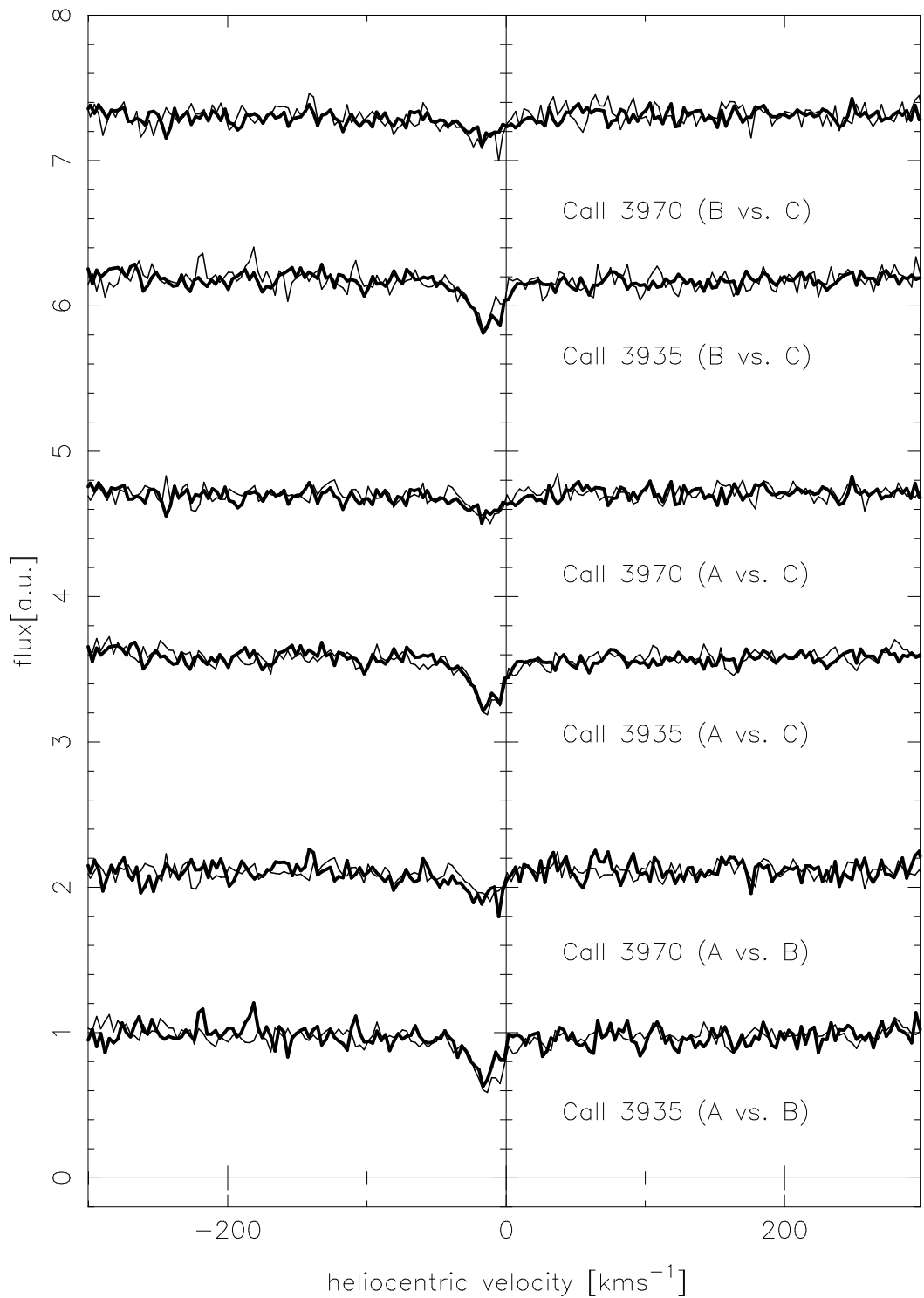}
\caption{\small 
absorption spectra of the CaII 3935, 3970
\AA\ absorption from the Milky Way (heliocentric velocities). The three
lines of sight A, B, and C, are compared in pairs. The uppermost three
spectra show the two transitions in lines of sight B (thin line) versus
C (thick line), the middle two spectra show A (thin line) versus C
(thick line), and the bottom two spectra A (thin line) versus B (thick
line), respectively.  Here and in the other figures the spectra are
normalized to a unit continuum and offset along the ordinate (labelled
in abitrary units) for better visibility.}
\end{figure}

\begin{figure}[t]
\figurenum{3}
\includegraphics*[scale=0.8,angle=-0.]{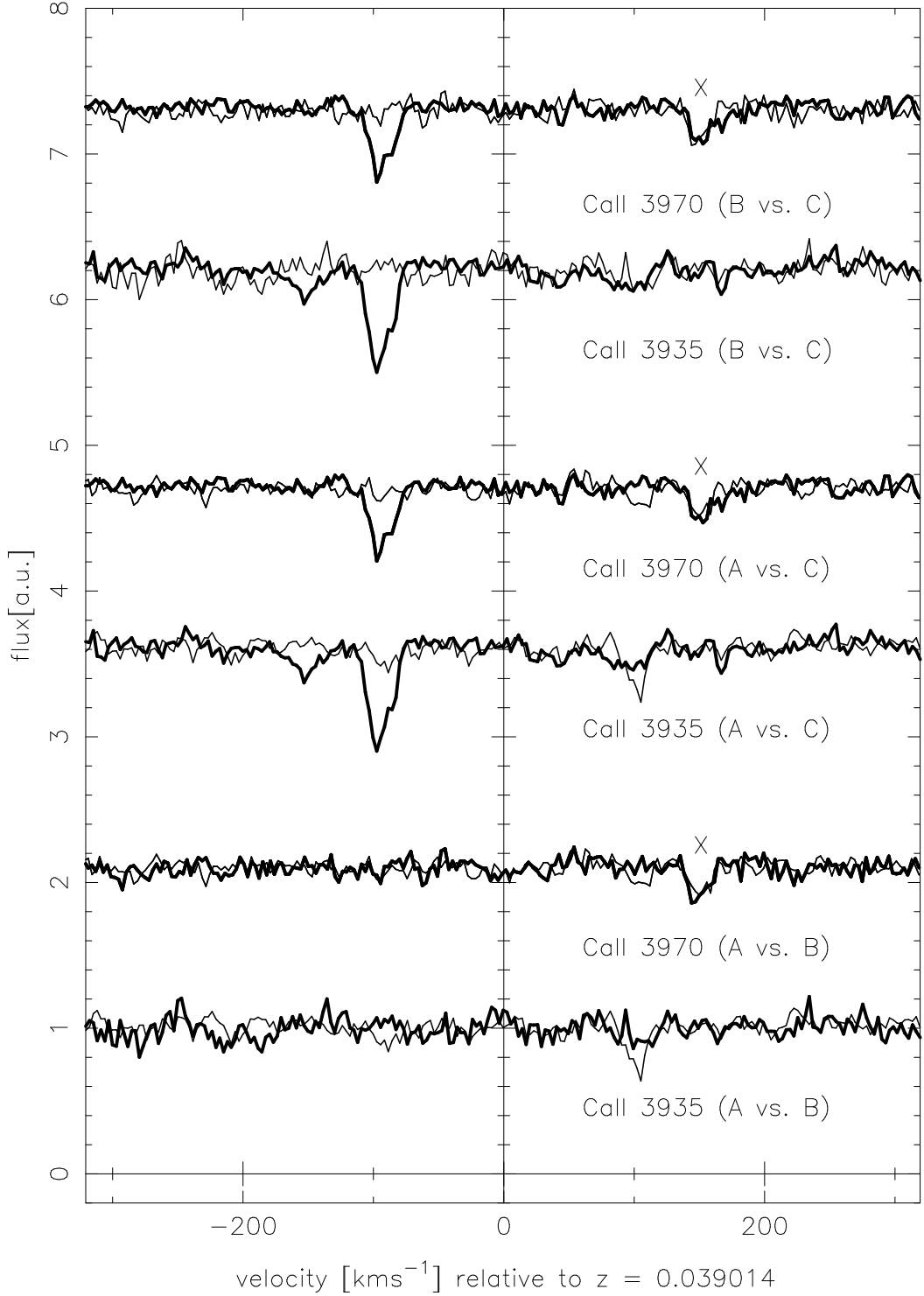}
\caption{\small 
spectra of the CaII 3935, 3970
\AA\ arising in the lensing galaxy (the origin of the
velocity scale corresponds to the galaxy redshift given by Foltz et al.
1992). The three lines of sight A, B, and C, are again compared in pairs. 
The lines marked with ``x'' are
interlopers of metal lines at different redshift. The proper beam
separation between A and B is here 1.89 $h_{50}^{-1}$ kpc, between A
and C 1.41 $h_{50}^{-1}$ kpc, and between B and C 1.45  $h_{50}^{-1}$
kpc.  }
\end{figure}

\begin{figure}[t]
\figurenum{4}
\includegraphics*[scale=0.85,angle=-0.]{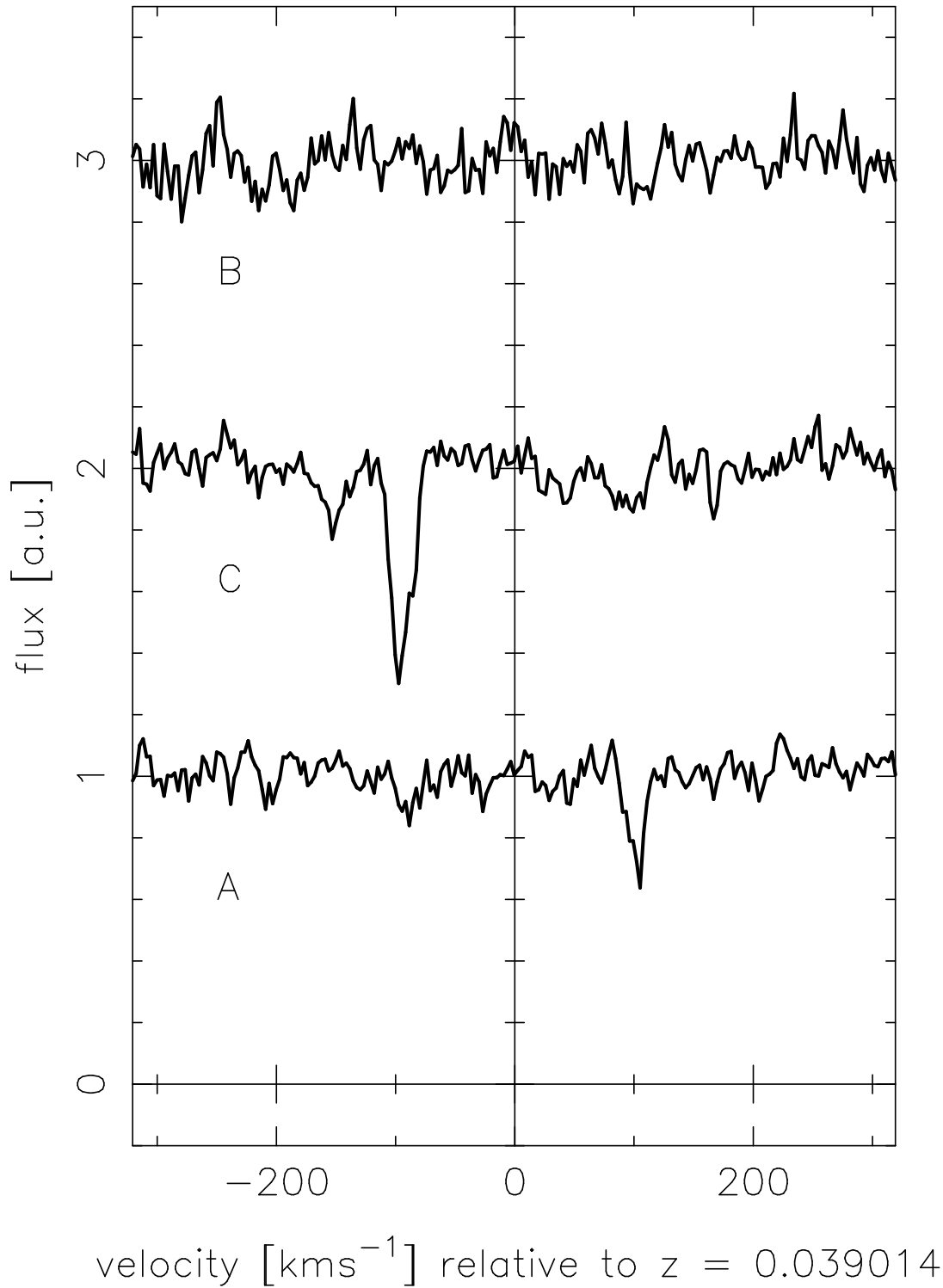}
\caption{\small 
Same system as in Fig. 3, (CaII at z=0.038) but now showing only the 3935 \AA\ transition
with the three systems overplotted offset by the height of the normalized continuum.
The three lines of sight A, C, and B, are arranged from bottom to top such that
the separation from image A increases as one goes from the bottom to the top spectrum.
 The proper beam separation between A and B is here 1.89 $h_{50}^{-1}$ kpc, between A and C 1.41 $h_{50}^{-1}$ kpc, and between B and C 1.45  $h_{50}^{-1}$ kpc.
}
\end{figure}

\begin{figure}[t]
\figurenum{5}
\includegraphics*[scale=0.8,angle=-0.]{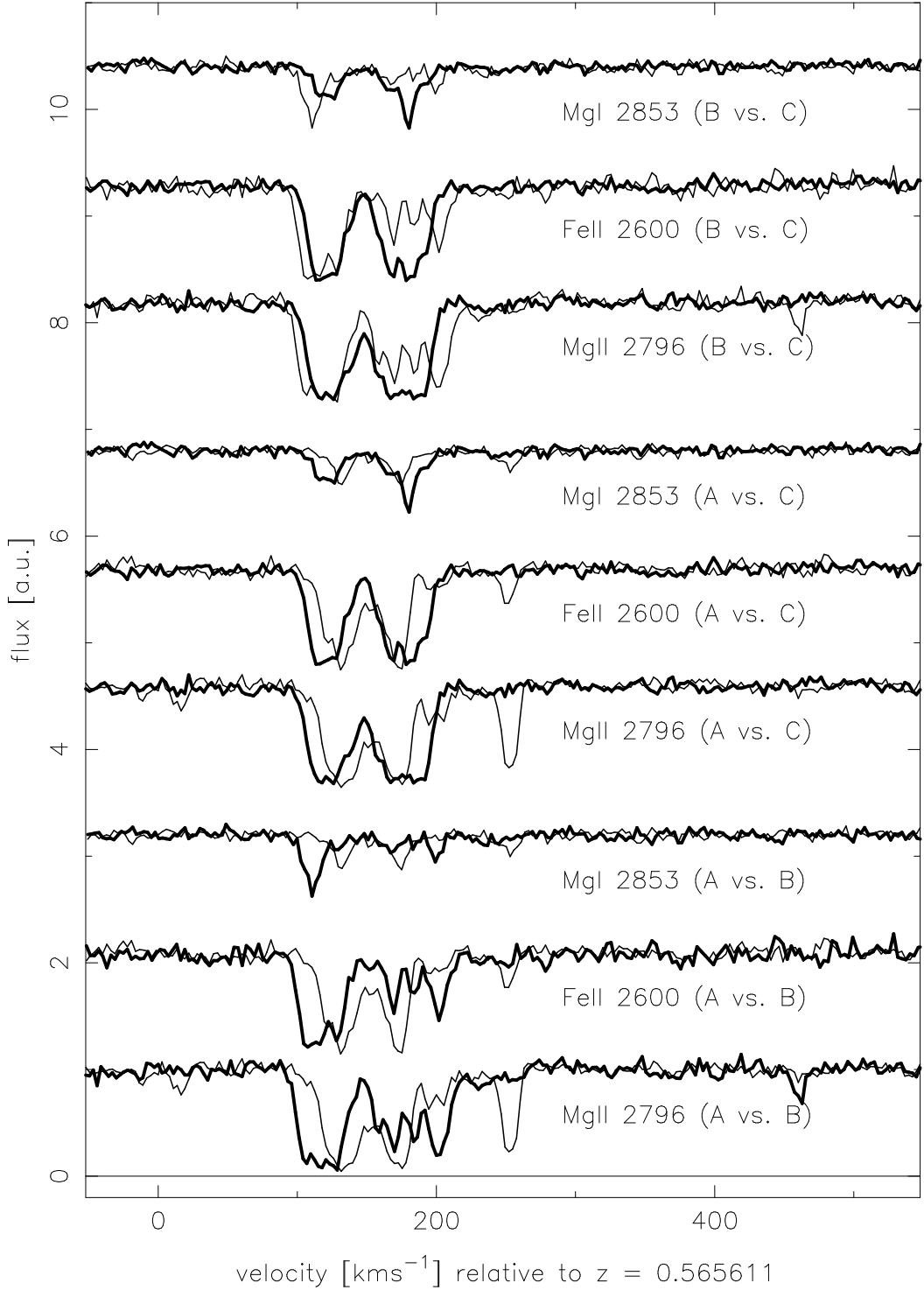}
\caption{\small 
absorption spectra of the three transitions MgI 2853,
FeII 2600, and MgII 2796 in the strong low ionization system near
z=0.5656. The three lines of sight A, B, and C, are compared in pairs.
The uppermost three spectra show the three transitions in lines of
sight B (thin line) versus C (thick line), the middle three spectra
show A (thin line) versus C (thick line), and the bottom three spectra
A (thin line) versus B (thick line), respectively. The proper beam
separation between A and B is here 0.66 $h_{50}^{-1}$ kpc, between A
and C 0.49 $h_{50}^{-1}$ kpc, and between B and C 0.50 $h_{50}^{-1}$
kpc.
}
\end{figure}

\begin{figure}[t]
\figurenum{6}
\includegraphics*[scale=0.65,angle=-90.]{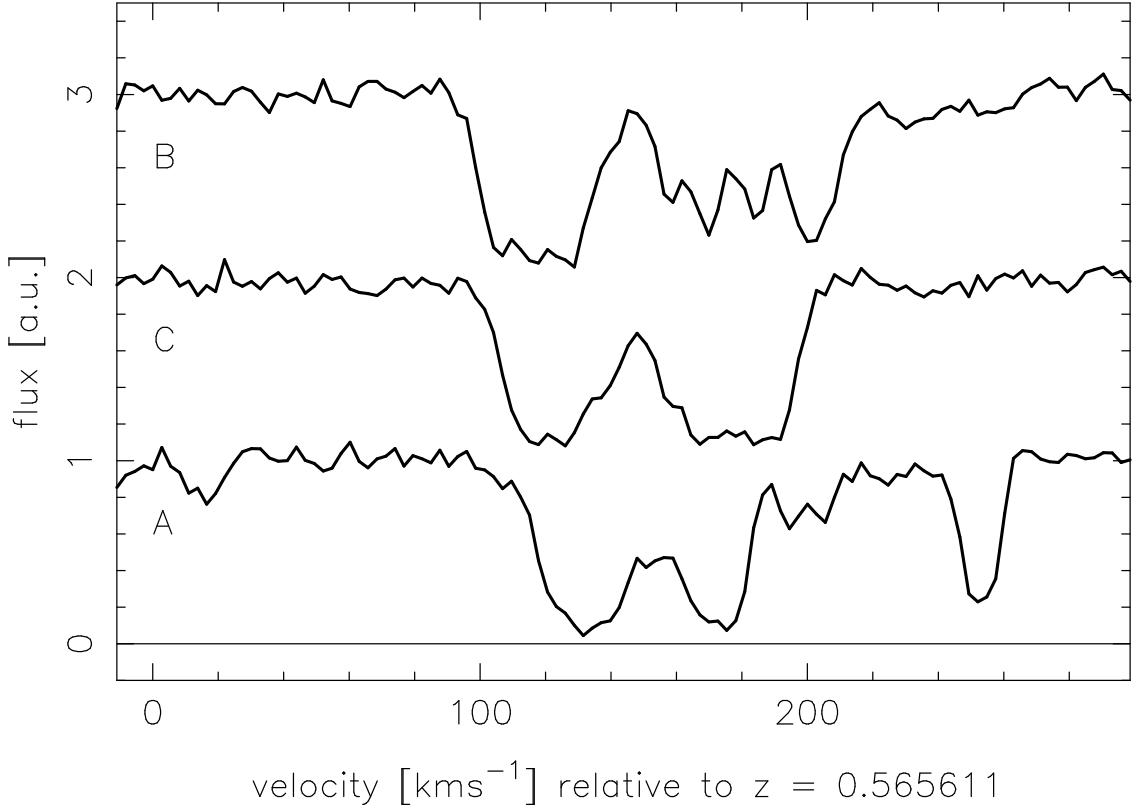}
\caption{\small 
absorption spectra of the transition MgII 2796 (at
z=0.56561), in the strong low ionization system shown already in the previous
figure. The three lines of sight A, B, and C are overplotted with offsets of
1.0 for clarity. There is a minimum of absorption near 150 kms$^{-1}$
on the arbitrary velocity scale. Also, going from the bottom to the top
spectrum the separation from the A image increases (moving in a
north-east direction on the sky - see Fig. 1) and the two stronger
groups separated by the 150 kms$^{-1}$ gap move further apart. The
proper beam separation between A and B is 0.66 $h_{50}^{-1}$ kpc,
between A and C 0.49 $h_{50}^{-1}$ kpc, and between B and C 0.50
$h_{50}^{-1}$ kpc.
}
\end{figure}

\begin{figure}[t]
\figurenum{7}
\includegraphics*[scale=0.8,angle=-0.]{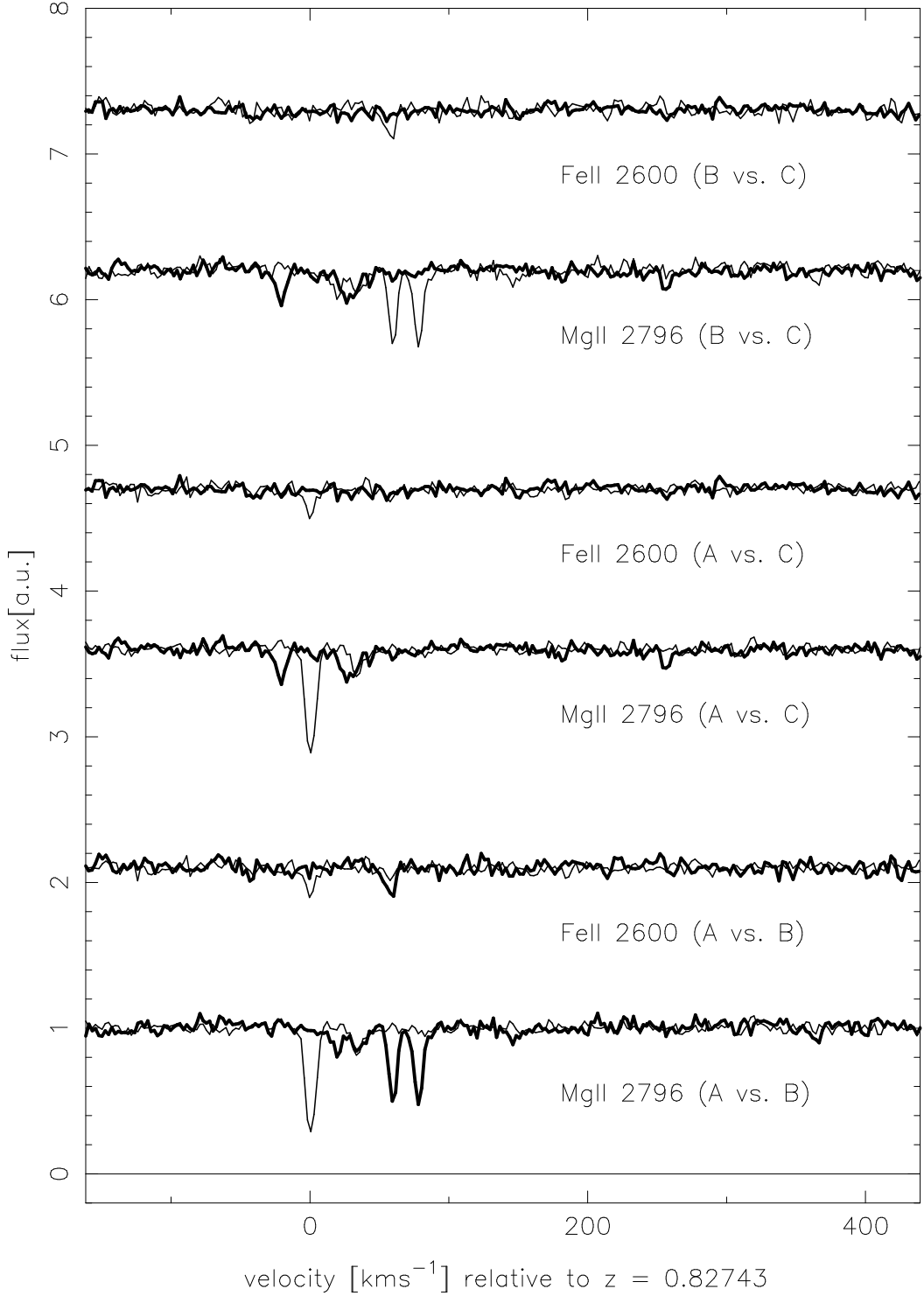}
\caption{\small 
absorption spectra of the two transitions  FeII 2600,
and MgII 2796 in the strong low ionization system near z=0.827. The
three lines of sight A, B, and C, are compared in pairs. The uppermost
two spectra show the two transitions in lines of sight B (thin line)
versus C (thick line), the middle two spectra show A (thin line) versus
C (thick line), and the bottom three spectra A (thin line) versus B
(thick line), respectively.  The proper beam separation between A and B
is  0.38 $h_{50}^{-1}$ kpc, between A and C 0.29 $h_{50}^{-1}$ kpc, and
between B and C 0.30 $h_{50}^{-1}$ kpc.
}
\end{figure}

\begin{figure}[t]
\figurenum{8}
\includegraphics*[scale=0.85,angle=-0.]{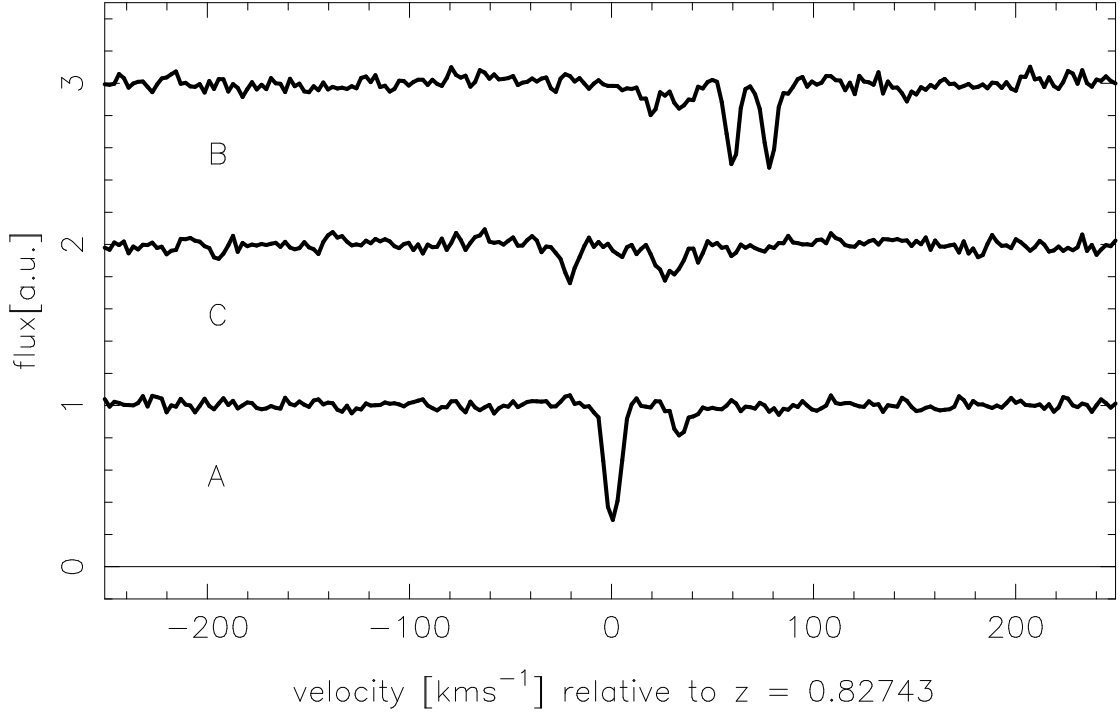}
\caption{\small 
Same system as in Fig. 7 (z=0.827), but now showing
only MgII 2796. All three systems are overplotted offset by the height
of the normalized continuum.  The three lines of sight A, C, and B, are
arranged from bottom to top such that the separation from image A
increases as one goes from the bottom to the top spectrum.  The proper
beam separation between A and B is  0.38 $h_{50}^{-1}$ kpc, between A
and C 0.29 $h_{50}^{-1}$ kpc, and between B and C 0.30 $h_{50}^{-1}$
kpc. }
\end{figure}

\begin{figure}[t]
\figurenum{9}
\includegraphics*[scale=0.85,angle=-0.]{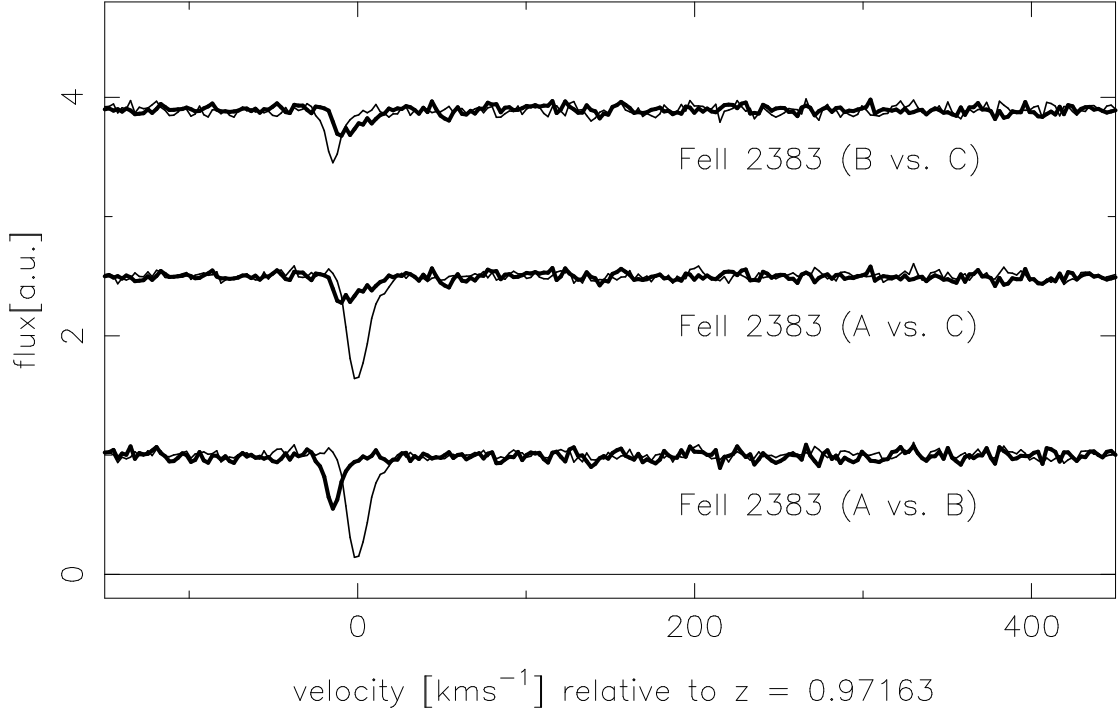}
\caption{\small 
absorption spectra of the  FeII 2383 transition low
ionization system near z=0.97163. The three lines of sight A, B, and C,
are compared in pairs. The uppermost  spectrum show the transition in
lines of sight B (thin line) versus C (thick line), the middle spectrum
show A (thin line) versus C (thick line), and the bottom spectrum A
(thin line) versus B (thick line), respectively.  The proper beam
separation between A and B is  0.28 $h_{50}^{-1}$ kpc, between A and C
0.21 $h_{50}^{-1}$ kpc, and between B and C 0.22 $h_{50}^{-1}$ kpc.}
\end{figure}

\begin{figure}[t]
\figurenum{10}
\includegraphics*[scale=0.85,angle=-0.]{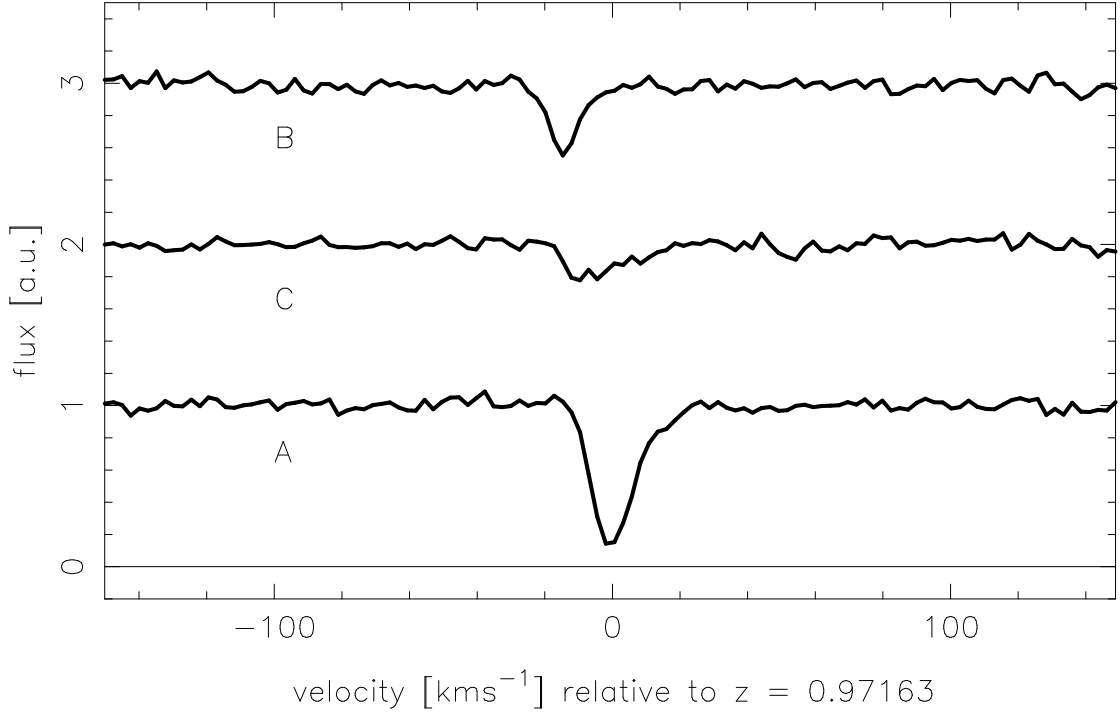}
\caption{\small 
Same system as in Fig. 9, (FeII 2353 at z=0.97163) but
now with the all three systems overplotted offset by the height of the
normalized continuum.  The three lines of sight A, C, and B, are
arranged from bottom to top such that the separation from image A
increases as one goes from the bottom to the top spectrum.  The proper
beam separation between A and B is  0.28 $h_{50}^{-1}$ kpc, between A
and C 0.21 $h_{50}^{-1}$ kpc, and between B and C 0.22 $h_{50}^{-1}$
kpc.}
\end{figure}

\begin{figure}[t]
\figurenum{11}
\includegraphics*[scale=0.65,angle=-90.]{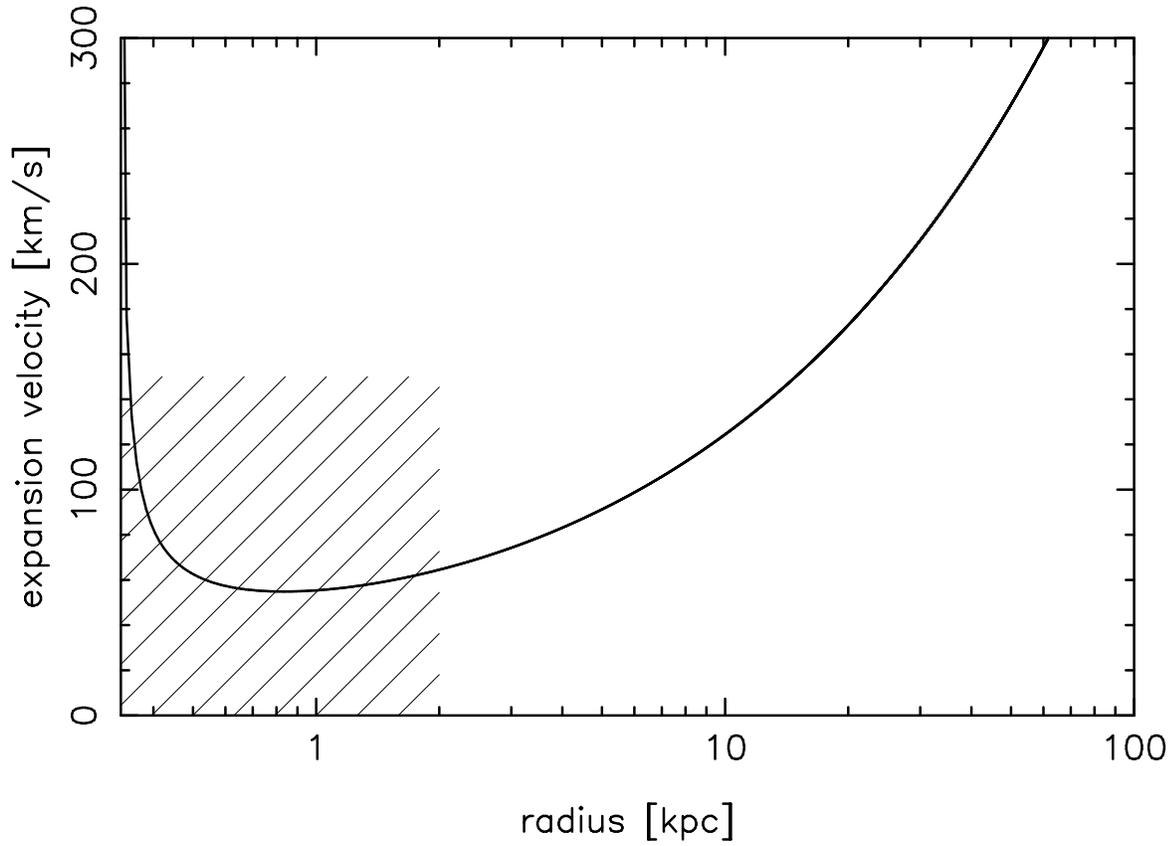}
\caption{\small 
expansion velocity of a spherical bubble consistent
with our observations, as a function of radius. The curve $v(R)$ was
computed from equation \ref{vee}. For comparison the hatched area shows
the approximate ranges of sizes and expansion velocities of HI shells
and supershells from the literature.}
\end{figure}

\begin{figure}[t]
\figurenum{12}
\includegraphics*[scale=0.5,angle=-0.]{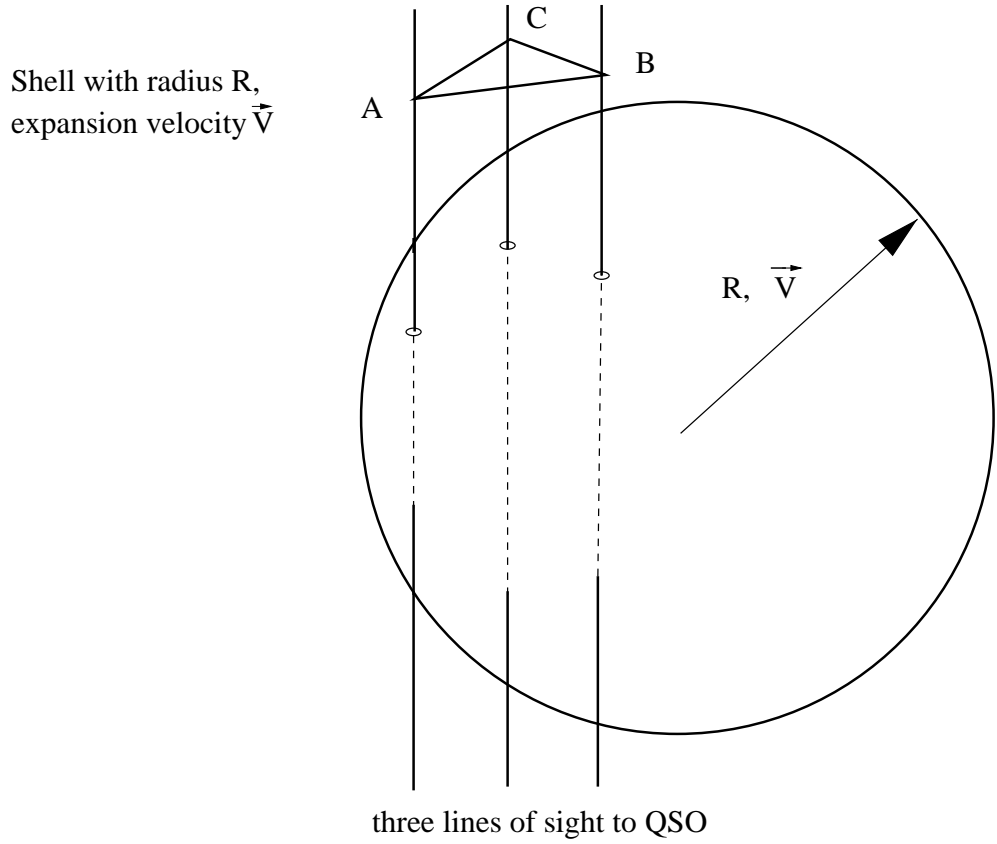}
\caption{\small 
schematic drawing of an expanding spherical bubble intersected
by three lines of sight.}
\end{figure}

\end{document}